\documentclass[a4paper,11pt]{article}
\pdfoutput=1
\usepackage{jcappub} 
\usepackage{slashed}
\usepackage{caption}
\allowdisplaybreaks
\title{\boldmath Recent Developments in Degenerate Higher Order Scalar Tensor Theories}

\author{Andrei Lazanu}

\affiliation{Department of Physics and Astronomy, University of Manchester, Manchester M13 9PL, United Kingdom}

\abstract{Degenerate Higher Order Scalar Tensor (DHOST) theories are the most general scalar-tensor theories whose Lagrangian depends on the metric tensor and a single scalar field and its derivatives up to second order. They propagate only one scalar degree of freedom, without being plagued by Ostrogradsky instabilities. This is achieved through certain degeneracies of the functions forming their Lagrangian. They generalise the Horndeski and beyond-Horndeski theories. Originally proposed to describe the late-time acceleration of the expansion of the universe, generalising the cosmological constant, they can also be used to build models of the early universe, to describe inflation or alternatives to standard inflation. In the late universe, they modify the standard Vainstein screening mechanism from Horndeski theories (which can have observable consequences) and are suited to build black hole models, featuring non-stealth Kerr black hole solutions. In this work their phenomenology is reviewed, looking at their basic properties, their parameterisations and classifications, focusing on solutions in the early and the late universe and at cosmological and astrophysical constraints.}

\begin{document}
\maketitle
\flushbottom

\section{Introduction}
The General Theory of Relativity (GR), proposed more than a century ago, supplemented by a cosmological constant, has been successfully used to describe the universe. Its predictions are in agreement with observations, including the accelerated expansion of the universe \cite{SupernovaSearchTeam:1998fmf, SupernovaCosmologyProject:1998vns} and the presence of gravitational waves \cite{LIGOScientific:2016aoc}.

The standard cosmological model, $\Lambda$CDM, consists of dark energy in the form of a cosmological constant ($\Lambda$), cold dark matter and ordinary matter. The cosmological constant, originally proposed by Einstein, leads to the so-called ``cosmological constant problem'', a discrepancy of 120 orders of magnitude between the value of the vacuum energy predicted by quantum field theory and the observed one from cosmology \cite{Carroll:1991mt}. In strong gravitational fields, such as in the vicinity of black hole singularities, predictions of GR are no longer accurate. These issues have led to a large number of alternative theories being developed, trying to relate the expansion of the universe to dark energy, which is also of an unknown nature, and also to find a way to parameterise it and to constrain it with various cosmological and astrophysical probes. The main alternatives are models of dark energy \cite{Frieman:2008sn} and modified gravity \cite{Clifton:2011jh}. Lovelock's theorem \cite{Lovelock:1971yv, Lovelock:1972vz} states that GR is the only 4D theory that has second-order Euler-Lagrange equations that can be build only from the metric. Hence, extensions or alternatives to GR must involve at least an additional degree of freedom. Various such models have been developed, including models depending on a single additional scalar field: 
Brans-Dicke \cite{Brans:1961sx}, quintessence \cite{Ratra:1987rm}, kinetic dominated scalar fields (k-essence) \cite{Chiba:1999ka,Armendariz-Picon:2000nqq,Armendariz-Picon:2000ulo}, chameleons \cite{Khoury:2003aq}, dilatons \cite{Brax:2010gi}, galileons \cite{Nicolis:2008in}, covariant galileons \cite{Deffayet:2009wt}. Other models have also been studied, see e.g. Refs. \cite{Nojiri:2017ncd, Shankaranarayanan:2022wbx} for reviews and the classification of models of modified gravity. A more general class of models is represented by Horndeski theories \cite{Horndeski1974, Horndeski:2024sjk}, of which the previous mentioned theories are special cases. These theories are characterised by second-order Euler-Lagrange equations, which was thought to be a requirement for the theories to be ghost-free (in order for the theory to be stable, one needs to avoid the generations of Ostrogradsky ghost instabilities \cite{Ostrogradsky:1850fid} which yield additional degrees of freedom).  Therefore, much of the interest has focused on theories which have second-order equations of motion. It turns out that this last requirement is not necessary, and thus they have been generalised to beyond Horndeski (Gleyzes-Langlois-Piazza-Vernizzi) \cite{Zumalacarregui:2013pma,Gleyzes:2014dya, Gleyzes:2014qga} and finally to the Degenerate Higher Order Scalar Tensor (DHOST) theories, first introduced in \cite{Langlois:2015cwa}. These are the most general scalar-tensor theories that propagate one scalar degree of freedom and which involve up to second-order derivatives of a single scalar field in their Lagrangians. They include all the scalar-tensor theories having these properties, including the Horndeski and beyond Horndeski ones.

In parallel, the origin and the first moments of the universe are also of great interest. Similarly to the late-time epoch previously described, the universe also underwent a brief period of accelerated expansion, known as inflation, just after the initial Big Bang. The theory describing it was developed from the late 1970s by several theoretical physicists including Alexei Starobinsky, Alan Guth, Stephen Hawking and Andrei Linde. The theory was originally built to solve various cosmological puzzles, such as the horizon problem, the flatness problem and the magnetic monopole problem. It has subsequently been shown that it can also be used to explain the presence of the structures that we observe in the universe today through the rapid growth of primordial quantum fluctuations \cite{Mukhanov:1981xt, Hawking:1982cz, Guth:1982ec, Starobinsky:1982ee}. This epoch has been constrained by cosmological probes such as \textit{Planck} \cite{Akrami:2018odb}. Its nature and the mechanisms which govern it remain nevertheless unknown and are active fields of research \cite{Turok:2002yq}. Many of the models describing inflation are based on the presence of at least a scalar field, the inflaton. Additionally, classical inflation usually assumes a `Big Bang' moment at the beginning of the universe, associated with a singularity. This assumption is relaxed in various alternative theories to inflation \cite{Gasperini:1992em,Battefeld:2005av,Creminelli:2010ba}.  The similar characteristics of the two epochs -- inflation and late-time accelerated expansion of the universe -- make the use of similar types of models appropriate and hence theories that are developed for one era can be adapted to the other \cite{Nojiri:2017ncd}.

The early developments of DHOST theories are reviewed in \cite{Langlois:2018dxi}. The paper presents the emergence of DHOST theories through the generalisation of Horndeski and beyond Horndeski ones and discusses their cosmological consequences as well as the gravitational properties of astrophysical bodies in this context. The mechanisms concerning the degeneracies are presented in detail, followed by a classification of the models and then the study of the theories within the formalism of the Effective Field Theory of dark energy and screening mechanisms. Results related to DHOST theories in the late universe and black hole solutions are also reviewed in \cite{Kobayashi:2019hrl}.

In this review, we discuss the phenomenological consequences of DHOST theories in cosmology in both phases of  accelerated expansion of the universe, and also for astrophysical objects. We introduce the DHOST theories in Section \ref{sec:theories}, and briefly mention the traditional classification of the quadratic ones. The Effective Field Theory of Dark Energy and its implications on DHOST models is then presented in Section \ref{sec:eft}. The consequences of these theories on astrophysical bodies (neutron stars and black holes) and the Vainshtein mechanism are the subject of Section \ref{sec:astro}. Other aspects, both cosmological and astrophysical, as presented in the literature, are 
mentioned  in Section \ref{sec:etc}. A DHOST model for inflation is presented in Section \ref{sec:inflation}, while an alternative model, DHOST Genesis, is reviewed in Section \ref{sec:bounce}. We conclude in Section \ref{sec:conclusions}. 

\section{DHOST Theories}
\label{sec:theories}
DHOST theories involve derivatives up to second order in a scalar field $\phi$ and can be classified in terms of the powers of derivatives present. So far, quadratic \cite{Langlois:2015cwa, Crisostomi:2016czh,Achour:2016rkg,deRham:2016wji} and cubic \cite{BenAchour:2016fzp} theories have been studied. Systematically classifying and analysing theories beyond third order becomes complicated due to the large number of terms involved. However, extensions to such theories have been studied phenomenologically. In Refs. \cite{Takahashi:2017pje, Langlois:2018jdg} a generalisation of the so-called mimetic gravity model has been discussed. More recently in Refs. \cite{Takahashi:2021ttd, Takahashi:2022mew,Takahashi:2023vva} another extension has been proposed by performing a higher-derivative generalisation of disformal transformations on the Horndeski action. The most general action containing up to cubic terms is given by

\begin{align}
\label{action}
S[g,\phi] &= \int d^4 x \, \sqrt{- g }
\left[ F_{(2)}(X,\phi) \,  {}^{(4)}R+P(X,\phi) + Q(X,\phi)  \square\phi + \sum_{i=1}^{5}A_i(X,\phi)  L^{(2)}_i
\right. \nonumber \\
 &+ \left.F_{(3)}(X, \phi) \, G_{\mu\nu} \phi^{\mu\nu}  +   \sum_{i=1}^{10} B_i(X,\phi) L^{(3)}_i  \right] \,,
\end{align}
where
\begin{equation}
\phi_\mu\equiv \nabla_\mu\phi\,, \qquad \phi_{\mu\nu}\equiv \nabla_\nu\nabla_\mu\phi\,, \qquad X\equiv \phi^\mu\phi_\mu\, ,
\end{equation}
and the elementary quadratic and cubic Lagrangians are

\begin{align}
& L^{(2)}_1 = \phi_{\mu \nu} \phi^{\mu \nu} \,, \qquad
L^{(2)}_2 =(\Box \phi)^2 \,, \qquad
L^{(2)}_3 = (\Box \phi) \phi^{\mu} \phi_{\mu \nu} \phi^{\nu} \,,  \label{Lag_quad1} \\
& L^{(2)}_4 =\phi^{\mu} \phi_{\mu \rho} \phi^{\rho \nu} \phi_{\nu} \,, \qquad
L^{(2)}_5= (\phi^{\mu} \phi_{\mu \nu} \phi^{\nu})^2\,, \label{Lag_quad2}
\end{align}
and 
\begin{align}
& L^{(3)}_1=  (\Box \phi)^3  \,, \quad
L^{(3)}_2 =  (\Box \phi)\, \phi_{\mu \nu} \phi^{\mu \nu} \,, \quad
L^{(3)}_3= \phi_{\mu \nu}\phi^{\nu \rho} \phi^{\mu}_{\rho} \,,   \\
& L^{(3)}_4= \left(\Box \phi\right)^2 \phi_{\mu} \phi^{\mu \nu} \phi_{\nu} \,, \quad
L^{(3)}_5 =  \Box \phi\, \phi_{\mu}  \phi^{\mu \nu} \phi_{\nu \rho} \phi^{\rho} \,, \quad
L^{(3)}_6 = \phi_{\mu \nu} \phi^{\mu \nu} \phi_{\rho} \phi^{\rho \sigma} \phi_{\sigma} \,,  \qquad  \\
& L^{(3)}_7 = \phi_{\mu} \phi^{\mu \nu} \phi_{\nu \rho} \phi^{\rho \sigma} \phi_{\sigma} \,, \quad
L^{(3)}_8 = \phi_{\mu}  \phi^{\mu \nu} \phi_{\nu \rho} \phi^{\rho}\, \phi_{\sigma} \phi^{\sigma \lambda} \phi_{\lambda} \,,  \qquad  \\
& L^{(3)}_9 = \Box \phi \left(\phi_{\mu} \phi^{\mu \nu} \phi_{\nu}\right)^2  \,, \quad
L^{(3)}_{10} = \left(\phi_{\mu} \phi^{\mu \nu} \phi_{\nu}\right)^3 \,.
\end{align}
These are all the possible contractions of the second-order derivatives $\phi_{\mu\nu}$ with the metric $g_{\mu\nu}$ and with the scalar field gradient $\phi_\mu$.   

The functions $A_i$ and $B_i$ must satisfy, however, certain degeneracy conditions such that the theory produces only a single scalar degree of freedom and no ghosts \cite{Langlois:2015cwa, Motohashi:2016ftl, Motohashi:2017eya, Motohashi:2018pxg}.

We note that the Horndeski theories are obtained by setting
\begin{align}
&F_{(2)}=G_4\,, \qquad A_1=-A_2=  2 G_{4,X}\,, \qquad A_3=A_4=A_5=0\,, \nonumber \\
&F_{(3)}=G_5 \,,\qquad 3B_1=-B_2=\frac32B_3=G_{5,X}\,, \qquad B_i=0\ (i=4,\dots, 10) \, .
\end{align}

The degeneracies arising in DHOST theories can be understood as follows. When considering beyond Horndeski and DHOST theories, the corresponding Euler-Lagrange equations can be of higher than second order. They can be equivalently reduced to a system of second-order equations by introducing additional constraint variables. Through a Hamiltonian formulation of this system, one can count the number of degrees of freedom of the system. If the matrix of coefficients is singular, however, then the number of degrees of freedom is reduced. This can be translated into conditions on the  functions appearing in the Lagrangian and it can be used to classify the DHOST theories. An intuitive explanation on the degeneracies appearing in DHOST theories is presented in Ref. \cite{Langlois:2015cwa}.

The degeneracy conditions place additional constraints on the functions $A_i$ and $B_i$. Based on whether the theories are quadratic, cubic or mixed, DHOST theories can be classified in several classes: quadratic theories (containing 7 subclasses), cubic theories (with 9 subclasses) and mixed quadratic and cubic theories (25 subclasses). The quadratic theories are presented in detail in  \cite{Langlois:2015cwa, Achour:2016rkg, Crisostomi:2016czh}, and the cubic ones in \cite{BenAchour:2016fzp}.  In the case of quadratic theories, a geometric interpretation of the degeneracies is discussed in Ref. \cite{Langlois:2020xbc}.

As these results are very technical, we will present in the following only the classification of quadratic theories, leaving the interested reader to discover the classification of the cubic and mixed quadratic/cubic theories in \cite{BenAchour:2016fzp}.

For quadratic theories, the degeneracy conditions are given by
\begin{equation}
D_0(X)=0, \qquad D_1(X)=0, \qquad D_2(X)=0\,,    
\end{equation}
where
\begin{align}
\label{D0}
D_0(X)&= -4 (A_1+A_2) \left[X F (2 A_2+X A_4+4 F_X)-2F^2-8X^2F_X^2\right]\,,\\
D_1(X)&= 4\left[X^2A_2 (3A_1+A_2)-2F^2-4X F A_1\right]A_4 +4 X^2 F(A_1+A_2)A_5  \nonumber\\
&+8XA_1^3-4(F+4X F_X-6X A_1) A_2^2 -16(F+5X F_X)A_1 A_2 \nonumber \\
&+4X(3F-4X F_X) A_2 A_3 
-X^2 F A_2^2 +32 F_X(F+2X F_X) A_1 \nonumber \\
&-16F F_X A_2-8F (F-X F_X)A_3+48F F_X^2 \,, \\
D_2(X)& = 4\left[ 2F^2+4X F A_1-X^2 A_2(3 A_1+A-2)\right] A_5  +4(2A_1-XA_3-4F_X)A_2^2 \nonumber \\
&+ 4 A_2^3+3X^2 A_2 A_3^2
-4X F A_3^2+8 (F+X F_X)A_2 A_3 -32 F_X A_1 A_2 \nonumber \\
&+16F_X^2 A_2+32 F_X^2 A_1-16 F F_X A_3\,.
\end{align}
Here we have dropped the subscript from the function $F_{(2)}$, as we are only dealing with quadratic theories. The classification of the theories is given in terms of how these three equations can be satisfied simultaneously. We express in square brackets the naming used in Ref. \cite{Crisostomi:2016czh}.

\subsection*{Class I  ($A_1+A_2=0$)}
\label{subsectionA}  
This class is characterised by
\begin{equation}
A_2=-A_1 \,.
\end{equation}
It has two sub-classes,

\begin{itemize}
    \item Class Ia [N-I] ($F-X A_1 \neq 0$)
    
There are three free functions, $F$, $A_1$ and $A_3$. The others are given by:
\begin{align}
\label{A4}
A_4&= \frac{1}{8(F - A_1 X)^2}\left[ 4 F \left(3 (A_1 - 2 F_X)^2 - 2 A_3 F\right) -A_3 X^2 (16 A_1 F_X + A_3 F)  \right. \nonumber \\
& \left. \qquad\qquad\qquad + 4 X \left(3 A_1 A_3 F + 16 A_1^2 F_X - 16 A_1 F_X^2 - 4 A_1^3 + 2 A_3 F F_X\right) \right] \,,  \\
\label{A5}
A_5 &= 
\frac{1}{8 (F - A_1 X)^2}(2 A_1 - A_3 X - 4 F_X) \left[A_1 (2 A_1 + 3 A_3 X - 4 F_X) - 4 A_3 F\right] \,.
\end{align}

    \item Class Ib [N-II] ($F-X A_1 = 0$)
    
There are three free functions, $F$, $A_4$ and $A_5$ and
\begin{equation}
A_3 =  \frac{2\left(F - 2 X F_{X}\right)}{X^2} \,.
\end{equation}
\end{itemize}

\subsection*{Class II ($F \neq 0$, $A_2 \neq - A_1$)}
In this class, $F \neq 0$ and $A_2 \neq A_1$. There are two subclasses:

\begin{itemize}
    \item Class IIa [N-IIIi] ($F-X A_1 \neq 0$))
    
There are three free functions, $F$, $A_1$ and $A_2$. Also,
\begin{align}
A_3 &= \frac{4 F_X (A_1 + 3 A_2)}{F} - \frac{2 (A_1 + 4 A_2 - 2 F_X)}{X} - \frac{4 F}{X^2} \,, \\
A_5 &= \frac{2}{F^2 X^3} \left[4 F^3 + F^2 X (3 A_1 + 8 A_2 - 12 F_X) \right. \nonumber \\
& \left. \qquad + 8 F\, F_X X^2 (F_X - A_1 - 3 A_2)+6 F_X^2 X^3 (A_1 + 3 A_2)\right] \,.
\end{align}

\item Class IIb [N-IIIii] ($F-X A_1 = 0$)

There are three free functions, $F$, $A_2$ and $A_3$ with $A_4$ and $A_5$ given by
\begin{align}
A_4 &=\frac{8 F_X^2}{F} - \frac{4 F_X}{X} \,, \\
A_5 &= \frac{1}{4 F X^3 (F + A_2 X)} \left[ F A_3^2 X^4 - 4 F^3 - 8 F^2 X (A_2 - 2 F_X)
 \right. \nonumber \\
& \left. - 4 F X^2 \left(4 F_X \left(F_X - 2 A_2\right) + A_3 F\right) + 8 F_X X^3 (A_3 F - 4 A_2 F_X) \right] \,.
\end{align}

\end{itemize}

\subsection*{Class III ($F=0$)}
This class corresponds to theories minimally coupled to gravity and can be split into three sub-classes.
\begin{itemize}
    \item Class IIIa [M-I] ($2A_1+A_4 X = 0$)
    
    There are three free functions $A_1$, $A_2$ and $A_3$, with $A_5$ given by
\begin{equation}
A_5 = \frac{4 A_1 \left( A_1 + 2 A_2 \right) - 4 A_1 A_3 X + 3 A_3^2 X^2}{4 \left(A_1 + 3 A_2\right) X^2}.
\label{MIa5}
\end{equation}
    \item Class IIIb [M-II] ($A_1+3A_2=0$)
    
There are three free functions $A_1$, $A_4$ and $A_5$, with $A_3$ given by
\begin{equation}
    A_3=\frac{2A_1}{3X} \, .
\end{equation}
\item Class IIIc [M-III] ($A_1=0$)

There are four free functions, $A_2$, $A_3$, $A_4$ and $A_5$.

\end{itemize}

We note that the classes are not necessary disjoint.

\section{Effective Field Theory of Dark Energy and Cosmological Constraints}
\label{sec:eft}

In order to confront predictions of a large number of dark energy and modified gravity theories to observations, an effective approach has been developed: the Effective Field Theory (EFT) of dark energy \cite{Gubitosi:2012hu, Gleyzes:2013ooa, Gleyzes:2014rba,Langlois:2017mxy}. This approach, which extends the EFT of inflation \cite{Creminelli:2006xe,Cheung:2007st}, is able to systematically treat all single-field models in the same framework, and hence to systematically constrain their parameters (see \cite{Frusciante:2019xia} for a review).

The metric is universally coupled to matter fields and the action is expressed in the most general unitary gauge consistent with the residual unbroken symmetries of spatial diffeomorphisms.
The background evolution only depends on three operators, which all the other operators starting at least at quadratic order in the perturbations. Hence, their effects can be studied independently. The main formalism and results relevant for DHOST theories are presented in the next paragraphs.

The framework is based on the ADM 3+1 decomposition of spacetime \cite{PhysRev.116.1322},
where spacetime is foliated into uniform scalar field hypersurfaces and expressed in terms of a lapse function $N$ and shift vector $N^i$,
\begin{equation}
ds^2=-N^2 dt^2 +h_{ij} (dx^i+N^i dt)(dx^j+N^jdt)\,.
\label{eq:ADM}
\end{equation}
Then the action can be expressed in terms of the extrinsic curvature tensor $K_{ij}$
\begin{equation}
K_{ij}\equiv \frac{1}{2N}\left(\dot h_{ij}-D_i N_j -D_jN_i\right)\,,
\end{equation}
as
\begin{equation}
S= \int d^3x \,  dt \,  N\sqrt{h}\,  L[N, K_{ij}, ^{(3)}R_{ij};t]\,,
\end{equation}
where $^{(3)}R_{ij}$ is the Ricci tensor associated with the spatial metric $h_{ij}$.

To investigate the dynamics of linear perturbations, one needs to consider the perturbations of the quantities in the action
\begin{equation}
\delta N\equiv N-\bar{N}\,, \qquad \delta K^i_j=K^i_j-H\delta^i_j\,, \qquad \delta^{(3)}R^i_j= ^{(3)}R^i_j\,,    
\end{equation}
where $H=\dot a/(\bar{N}a)$ is the Hubble parameter, and $^{(3)}R^i_j$ is a perturbation as $^{(3)}\bar{R}^i_j=0$.
The Lagrangian at quadratic order in DHOST theories can be expressed as
\begin{eqnarray}
\label{Squad}
 S_{\rm quad} &=& \int d^3x \,  dt \,  a^3  \frac{M^2}2\bigg\{ \delta K_{ij }\delta K^{ij}- \left(1+\frac{2}{3}\alpha_L\right)\delta K^2  +(1+\alpha_T) \bigg(\delta  ^{(3)}R\,  \frac{\delta \sqrt{h}}{a^3} + \delta_2 ^{(3)}R 
 \bigg)  
 \cr
&&\qquad \qquad \qquad \qquad 
 + H^2\alpha_K \delta N^2  +4 H \alpha_B \delta K \delta N+ ({1+\alpha_H}) \, ^{(3)}R  \, \delta N   
\cr
&&
\qquad \qquad \qquad \qquad 
+  4 \beta_1\,   \delta K  {\delta \dot N }   + \beta_2 \, {\delta \dot N}^2 +  \frac{\beta_3}{a^2}(\partial_i \delta N )^2   
\bigg\} \,, \qquad\quad 
\end{eqnarray}
where $\delta_2 ^{(3)}R$ represents the second order term in the perturbative expansion of $^{(3)}R$, and $M$, $\alpha_L$, $\alpha_T$, $\alpha_K$, $\alpha_B$, $\alpha_H$, $\beta_1$, $\beta_2$ and $\beta_3$ are time-dependent functions. 

Parameters  $M$,  $\alpha_T$, $\alpha_K$, $\alpha_B$ are specific to Horndeski theories, while the additional parameter $\alpha_H$ appears in beyond-Horndeski ones. The physical description of these parameters is discussed in \cite{Bellini:2014fua}. For quadratic and cubic theories, these EFT parameters can be expressed in terms of the $A_i$ and $B_i$ parameters as \cite{Langlois:2017mxy}

\begin{align}
{M^2}=&\ 2 F_2 - X[ 2 A_1+F_{3 \phi} - 6 (B_2+B_3)H\mu^2 ] \,, \label{eq:m2}\\ 
\alpha_T= & -1 +\frac{2F_2+X F_{3\phi}}{M^2}\;, \label{eq:at}
\\
\alpha_H=&-1 +\frac{1}{M^2}  \left[ 2F_2-X\left(4  F_{2X}+ 2F_{3X}H\mu^2 +F_{3\phi}\right)\right]\,, \\  
\alpha_L   = & \ \frac{3 X}{M^2} \left[   A_1+ A_2  - (9B_1+5 B_2+ 3 B_3)H\mu^2  \right] \;, \\
\beta_1=& \  \frac{X}{2 M^2} \left\{ 4  F_{2X}+2 A_2 +X A_3 +2  \left[2F_{3X}-9B_1-B_2 -X (3B_4+B_6)\right] H\mu^2 \right\} \,,
\\
\beta_2=&  -  \frac{2 X}{M^2}  \left\{ A_1+A_2 + X(A_3+A_4)+ X^2 A_5 
- 3 \left[3B_1+B_2 + X(2B_4+B_5)+ X^2  B_9)\right] H\mu^2 \right\} \,,\\
\beta_3=& - \frac{2 X}{M^2}  \left\{ 4  F_{2X}-2A_1- X A_4 + \left[4 F_{3X}+6B_2+3B_3+ X (3B_5+B_7)\right] H\mu^2 \right\}\, ,
\end{align}
where a linear assumption $\phi=\mu^2t$ is made.

Out of the four DHOST-specific parameters, only one is independent, and this is due to the degeneracy conditions. Hence, the DHOST theories can be classified as:
\begin{align}
\mathcal{C}_{\rm I}&:\quad \alpha_L=0\,, \qquad \beta_2=-6\beta_1^2\,,\qquad   \beta_3=-2\beta_1\left[2(1+\alpha_H)+\beta_1 (1+\alpha_T)\right]\,, \\
\mathcal{C}_{\rm II}&:\quad \beta_1=- (1+\alpha_L)\frac{1+\alpha_H}{1+\alpha_T}\,, \quad \beta_2=-6(1+\alpha_L) \frac{(1+\alpha_H)^2 }{(1+\alpha_T)^2}\,,\quad \beta_3=2\frac{(1+\alpha_H)^2}{1+\alpha_T}\,.
\end{align}

In order to investigate the stability of the solutions, one can derive the quadratic tensor action
\begin{equation}
\label{S_quad_T}
 S_{\rm quad, tensor} = \int d^3x \,  dt \,  a^3  \bigg\{\frac{M^2}8\left[\dot\gamma_{ij}^2-\frac{1+\alpha_T}{a^2}(\partial_k\gamma_{ij})^2\right]\bigg\} \,.
\end{equation}
Hence, the parameter $\alpha_T$ is associated to the tensor propagating velocity,
\begin{equation}
c_T^2=1+\alpha_T \,.
\label{c_T}
\end{equation}
In particular, from Eqs. \eqref{eq:m2} and \eqref{eq:at} one can see that the condition $\alpha_T=0$ corresponds to $A_1=0$.
For the scalar, the quadratic action becomes
\begin{equation}   
\label{actzIa}
S_{\rm quad, scalar}= \int d^3 x \, dt \, a^3   \frac{M^2 }{2} \bigg[  {A}_{\tilde \zeta}   \dot{\tilde \zeta}^2 -  {B}_{\tilde \zeta} \frac{(\partial_i \tilde \zeta)^2}{a^2}   \bigg] \, .
\end{equation}
For $\mathcal{C}_{\rm I}$ theories, $A_{\tilde \zeta}$ and  ${B}_{\tilde \zeta}$  are given by
\begin{align}
{A}_{\tilde \zeta} & = \frac{  1}{(1+\alpha_B-\dot \beta_1 /H)^2}  \bigg[\alpha_K+ 6\alpha_B^2 - \frac{6}{a^3 H^2 M^2} \frac{d}{dt} \left( a^3 H M^2 \alpha_B \beta_1 \right) \bigg]   \; , \label{AzI}\\
{B}_{\tilde \zeta} & =  -2 (1+\alpha_T)+\frac{2}{a M^2 }\frac{d}{dt}\bigg[\frac{a M^2 \big( 1+\alpha_H+\beta_1(1+\alpha_T)\big)}{H(1+\alpha_B)-\dot \beta_1}\bigg]   \;. \label{BzI} 
\end{align}
The stability conditions require
\begin{equation}
{A}_{\tilde \zeta}>0\,, \qquad {B}_{\tilde \zeta}>0\,.
\label{eq:stab}
\end{equation}
In the case of $\mathcal{C}_{\rm II}$ theories, ${B}_{\tilde \zeta}=-2(1+\alpha_T)$, which combined with the result for the tensor modes, shows that scalar and tensor modes cannot be stable simultaneously, and hence this class of theories is not viable.

The effects of the gravitational wave event GW170817 \cite{LIGOScientific:2017vwq}, observed by the LIGO/Virgo collaboration, and of the gamma-ray burst GRB 170817A \cite{LIGOScientific:2017zic} on the DHOST models have been discussed in  \cite{Langlois:2017dyl,Bordin:2020fww}. Fixing the gravitational wave speed to be equal to the speed of light only allows a Lagrangian of the form
\begin{align}
\label{DHOST}
& L^{_{\rm DHOST}}_{c_g=1}=  P + Q  \Box\phi +  F  \, {}^{(4)}\! R +  A_3\phi^\mu \phi^\nu \phi_{\mu \nu} \Box \phi  \nonumber\\
&+\frac{1}{8F} \bigg(48 F_X^2 -8(F-X F_X) A_3-X^2 A_3^2 \bigg) \phi^\mu \phi_{\mu \nu} \phi_\lambda \phi^{\lambda \nu} \nonumber\\
&+\frac{1}{2 F}\left(4F_X+X A_3\right) A_3(\phi_\mu \phi^{\mu \nu } \phi_\nu)^2 \;.
\end{align}
This restriction is based on the assumption that scales probed by the LIGO-Virgo measurements (which are of the order of  $10^3$ km), can be extrapolated by several orders of magnitude to cosmological scales. We note that, if the decay rate is too large, we would not have observed gravitational waves. In Ref. \cite{Creminelli:2018xsv}, the authors found that imposing $A_3 = 0$ on top of $A_1 = 0$ leads to a suppression of the decay rate of gravitational waves such that the decay time can be longer than the age of the Universe.

The cosmological evolution of these models has been discussed in \cite{Crisostomi:2017pjs, Crisostomi:2018bsp, Frusciante:2018tvu, Arai:2019zul, Boumaza:2020klg}. Some of the studies include a relaxation of the assumption related to the tensor mode speed, allowing much more general DHOST theories. 

We illustrate in this section the cosmological evolution of quadratic DHOST theories, as described in Ref. \cite{Boumaza:2020klg}. In this work, the gravitational wave speed constraints are not used, as no extrapolation from astrophysical to cosmological scales is assumed. The aim is to find self-accelerating solutions, corresponding to the late-time evolution of the universe and the transition between the matter and accelerated expansion epochs. 

The starting point is a homogeneous background, depending on a lapse function $N$ and a scale factor $a$,
\begin{equation}
ds^{2}=-N^2(t) dt^{2}+a^2(t)\delta_{ij}dx^{j}dx^{i} \,,
\end{equation}

The homogeneous action becomes:
\begin{eqnarray}
S_{\rm hom}[N,a,\phi] & = & \int dt\,a^{3}N\Biggl\{P+Q\left(\frac{\dot{N}}{N^{3}}\dot{\phi}-\frac{3\dot{a}}{a\,N^{2}}\dot{\phi}-\frac{\ddot{\phi}}{N^{2}}\right)-F_{\phi} \frac{6\dot{a}}{a\,N^{2}}\dot{\phi}\nonumber \\
 &  & \qquad \qquad -\frac{6f_{1}}{N^{2}}\left(\frac{\dot{a}}{a}+\frac{f_{2}}{4f_{1}}\left(\frac{\dot{N}\dot{\phi}{}^{2}}{N^{3}}-\frac{\ddot{\phi}\dot{\phi}}{N^{2}}\right)\right)^{2}\Biggr\} \, , \label{Shom}
\end{eqnarray}
where
\begin{eqnarray}
\label{f1f2}
f_{1}  \equiv  F- X A_{1}\,,\qquad
f_{2}  \equiv  4F_{X}- 2 A_{1}+XA_{3}\,,
\end{eqnarray}
and
\begin{equation}
X=-\frac{\dot{\phi}{}^{2}}{N^{2}}\,.
\end{equation}
The equations of motion from \eqref{Shom} appear to be higher than second order, but this is not the case due to the degeneracy conditions. This can be seen by defining an auxiliary scale factor $b$, 

\begin{equation}
a\equiv \Lambda\left(X,\phi\right)b\equiv e^{\lambda(X,\phi)} b\,. 
\end{equation}
Here $\lambda$ satisfies 
\begin{equation}
\lambda_{X}=-\frac{f_{2}}{8f_{1}}\,,
\label{lambdaX}
\end{equation}
and hence the quadratic terms in $\ddot \phi$ in the action (\ref{Shom}) are reabsorbed in the derivatives of the new scale factor. 

The Hubble parameter associated with this auxiliary scale factor is defined as
\begin{equation}
\label{Hb}
H_b\equiv \frac{\dot b}{N b}=H-\lambda_X \frac{\dot X}{N} -\lambda_\phi\frac{\dot\phi}{N}\,, \qquad
\dot{X}= \frac{2}{N^2} \left(\frac{\dot N \dot \phi^2}{N} - {\dot\phi \ddot\phi }\right) \, .
\end{equation}

From action, \eqref{Shom} one can express the Lagrangian in terms of the new expansion factor $b$ as
\begin{eqnarray}
L_{\rm hom} & = & \Lambda^{3}b^{3}N\Biggl\{-3  \lambda_{\phi}\frac{\dot{\phi}{}^{2}}{N^{2}}(2 f_1 \lambda_\phi + Q + 2 F_\phi)+P-6f_{1}H_{b}^{2} -\frac{3\dot{\phi}}{N}\left(4\lambda_{\phi} f_{1}+2F_{\phi}+Q\right)H_{b} \nonumber \\
 &  & \qquad +\left(Q-6\lambda_{X}\left(2F_{\phi}+Q\right)\frac{\dot{\phi}2}{N^{2}}\right)\left(\frac{\dot{N}\dot{\phi}}{N^{3}}-\frac{\ddot{\phi}}{N^{2}}\right)\Biggr\}.
\end{eqnarray}

The total Lagrangian $L$ is then the sum of the preceding one and the matter Lagrangian $L_m$.
The equations of motion are obtained by considering the Euler-Lagrange equations for $N$, $b$ and $\phi$. They yield generalised Friedmann equations and the scalar field equation of motion respectively,
\begin{eqnarray}
-\frac{d^{2}}{dt^{2}}\frac{\partial L}{\partial\ddot{\phi}}+\frac{d}{dt}\frac{\partial L}{\partial\dot{\phi}}-\frac{\partial L}{\partial\phi} & = & 0.
\end{eqnarray}
Then one can fix the time coordinate such that $N=1$ (and $\dot{N}=0$).

The matter is assumed to be described by a perfect fluid whose equation of state is $P=w\rho$, where $w$ is constant. The energy-momentum tensor of the fluid is defined as
\begin{equation}
T^{\mu\nu}=\frac{2}{\sqrt{-g}}\frac{\delta S_m}{\delta g_{\mu\nu}}\,,
\qquad S_m = \int d^4x \sqrt{-g} \, L_m \, .
\end{equation}
Then the variation of the matter Lagrangian is given by

\begin{equation}
\delta L_m= - a^3 \rho_m \delta N+ 3 N a^2 P_m \delta a\,,
\end{equation}
where $\rho_m$ and $P_m$ are  the fluid energy density and pressure, respectively.

The two Friedmann equations take the form
\begin{align}
 g_{0}+g_{1}H_{b}\dot\varphi+g_{2}H_{b}^{2} & =  \left(1-6 w \lambda_{X}\dot{\varphi}^{2}\right)  \rho_{m}  \, , \label{frw1}\\
 g_{3}+g_{4}(2\dot{H}_{b}+3H_{b}^{2})+g_{5}H_{b}\dot{\varphi}+g_{6}\ddot{\varphi}+g_{7}H_{b}\dot{\varphi} \ddot{\varphi} & =  -w \rho_{m} \,. \label{frw2}
\end{align}
and the scalar field equation can be written as
\begin{equation}
\frac{d}{dt}\left(b^{3}\Lambda^{3}J\right)+b^{3}\Lambda^{3} U=0\,,
\label{SFE}
\end{equation}
where 
\begin{eqnarray}
\label{UandJ}
U  =  g_{8}+g_{9}H_{b}\dot{\varphi}+g_{10}H_{b}^{2}+g_{11}\ddot{\varphi}\, , \qquad
J  =  g_{12}\dot{\varphi}+g_{13}H_{b}+g_{14}H_{b}^{2}\dot{\varphi} \,.
\end{eqnarray}
The coefficients $g_i$ can be expressed in terms of quantities in the Lagrangian and of $\lambda$ and are given in Appendix A of Ref. \cite{Boumaza:2020klg}.

To investigate the stability of the solutions of these equations, one needs to consider linear perturbations, and this has been described in the previous paragraphs. In the absence of matter, the stability conditions are given in \eqref{eq:stab}. If one also considers matter, then the analysis is significantly more complicated because there is an additional degree of freedom, $\delta \sigma$. Then the quadratic action can be expressed as

\begin{equation}
S_{\rm quad}=\int d^3x \, dt \, a^3 \frac{M^2}{2} \left[  \dot V^T \, {\bf K} \, \dot V - \frac{1}{a^2} \partial_i V^T \, {\bf G} \, \partial^i V + \dots \right] \,,
\end{equation}
where the vector $V^T=(\zeta,H \frac{\delta\sigma}{\dot\sigma})$  contains the two scalar degrees of freedom and the dots stand for the terms with fewer than two (space or time) derivatives, which are not relevant for the stability discussion. The kinetic and gradient matrices read  \cite{Crisostomi:2018bsp}
\begin{align}
{\bf K} &= \left(
\begin{matrix}
    A_\zeta + \frac{\rho_m(1+w_m)}{M^2 c_m^2 \left( H(1+\alpha_B) - \dot \beta_1\right)^2}    \qquad   &  \frac{ \rho_m (1+w_m) \left( 3 c_m^2 \beta_1 -1 \right)}{M^2 c_m^2 H\left( H(1+\alpha_B) - \dot \beta_1\right)} \\[4ex]
    \frac{ \rho_m (1+w_m) \left( 3 c_m^2 \beta_1 -1 \right)}{M^2 c_m^2 H \left( H(1+\alpha_B) - \dot \beta_1\right)}       & \frac{\rho_m(1+w_m)}{M^2 c_m^2 H^2}
\end{matrix} \right) \,,  \\[4ex]
{\bf G} &= \left(
\begin{matrix}
   B_\zeta    &  -\frac{ \rho_m (1+w_m) \left( 1 + \alpha_H + (1+ \alpha_T) \beta_1\right)}{M^2  H\left( H(1+\alpha_B) - \dot \beta_1\right)} \\[4ex]
    -\frac{ \rho_m (1+w_m) \left( 1 + \alpha_H + (1+ \alpha_T) \beta_1\right)}{M^2 H \left( H(1+\alpha_B) - \dot \beta_1\right)}     &  \frac{\rho_m(1+w_m)}{M^2H^2}
\end{matrix} \right)  \,. 
\end{align}
To avoid ghost and gradient instabilities, both matrices $\bf K$ and $\bf G$ must be positive definite. 
If the matter satisfies $c_m \ll 1$ and $w_m \ll 1$, one can expand the expressions of the eigenvalues of $\bf K$ and $\bf G$ with respect to $c_m$ and $w_m$. At leading order, the eigenvalues of $\bf K$ and $\bf G$, $\lambda_{K_{1,2}}$ and $\lambda_{G_\pm}$ respectively, satisfy,
\begin{equation}
\begin{split}
\lambda_{K_{1}}  & =  \frac{A_{\zeta}M^{2}H^{2}(1+\alpha_{B}-\beta_{1}')^{2}+6\rho_{m}\beta_{1}}{M^{2}H^{2}(1+(1+\alpha_{B}-\beta_{1}')^{2})}\,, \quad \;
\lambda_{K_{2}} = \frac{\rho_{m}}{c_{m}^{2}H^{2}M^{2}}\left[\frac{1}{(1+\alpha_{B}-\beta_{1}')^{2}}+1\right],\\
\lambda_{G_{\pm}} & = 
 \frac{B_{\zeta}}{2}+\frac{1}{2M^{2}}\left[\frac{\rho_{m}}{H^{2}}\pm\sqrt{\frac{4\rho_{m}^{2}(1+\alpha_{H}+(1+\alpha_{T})\beta_{1})^{2}}{H^{4}(1+\alpha_{B}-\beta_{1}')^{2}}+(\frac{\rho_{m}}{H^{2}}
 -M^{2}B_{\zeta})^{2}}\, \right]\,.
\end{split}
\end{equation}
Hence, $\lambda_{K_{2}}$ is always positive  while the signs of the three other eigenvalues depend on the specific background solution. 

In \cite{Boumaza:2020klg}, specific numerical examples are investigated, explicitly showing that there are stable DHOST theories.

Another possible instability in such theories is the Dolgov-Kawasaki  instability \cite{Dolgov:2003px}, which arises when a gravity theory allows for a dynamical equation for the scalar curvature through the trace of the metric field equations and the associated square mass is non-positive. In the case of Horndeski this has been shown not to be an issue \cite{Gomes:2020lvs}.

The EFT method has been used, together with the CMB likelihoods from \textit{Planck}, to place constraints on the EFT parameters in DHOST models \cite{Hiramatsu:2020fcd,Hiramatsu:2022fgn}. This work has relied upon the development of a Boltzmann solver that incorporates the EFT parameters and the use of a Markov-Chain Monte-Carlo method. Given the nature of the work, particular subclasses of DHOST have been chosen.
For a simple model parameterised in terms of the dark energy density $\Omega_{\rm DE}$ as $\alpha_{\rm K} = \Omega_{\rm DE}(t)/\Omega_{\rm DE}(t_0)$, $\alpha_{\rm B}=\alpha_{\rm T}=\alpha_{\rm M}=\alpha_{\rm H}=0$ and $\beta_1=\beta_{1,0}\Omega_{\rm DE}(t)/\Omega_{\rm DE}(t_0)$ and a $\Lambda$CDM background, a constraint $\beta_{1,0}=0.032_{-0.016}^{+0.013}$ (at 68\% CL) has been derived.
For a theory given by $\mathcal{L}_{\rm DHOST} = X + c_3X\Box\phi/\Lambda^3+ (M_{\rm Pl}^2/2+c_4X^2/\Lambda^6)R + 48c_4^2X^2/(M_{\rm Pl}^2 \Lambda^{12}+2c_4\Lambda^6X^2)\phi^\mu\phi_{\mu\rho}\phi^{\rho\nu}\phi_\nu$
 with two positive constant parameters, $c_3$ and $c_4$, the authors have obtained $c_3 = 1.59^{+0.26}_{-0.28}$ and the upper bound on $c_4$, $c_4<0.0088$ (at 68\% CL).

\section{Astrophysical Bodies and the Vainshtein Mechanism}
\label{sec:astro}
Significant work has been performed on studying the effects of DHOST theories on astrophysical objects, in particular gravitational screening, neutron stars and black holes. 

\subsection{The Vainshtein Mechanism}
\label{sec:vain}
Dark energy and modified gravity models can change the spacetime evolution compared to GR, but these modifications should not appear at small scales, where precise laboratory and solar system tests can be performed. This can be achieved by the so-called screening mechanisms. In the case of DHOST theories, the Vainshtein screening \cite{Vainshtein:1972sx} is at play. It has been shown to work well for Horndeski theories \cite{Kimura:2011dc,Narikawa:2013pjr,Koyama:2013paa}, whereas in the case of beyond-Horndeski theories the standard gravity is modified inside matter \cite{Kobayashi:2014ida}. This new behaviour has been studied for various astrophysical objects, as presented in  \cite{Sakstein:2017xjx}. For DHOST theories, this topic is discussed in general in \cite{Langlois:2017dyl, Crisostomi:2017lbg, Dima:2017pwp}, with specific examples in \cite{Hirano:2019scf,Laudato:2021mnm,Laudato:2022vmq,Laudato:2022drz}.

To illustrate how the Vainshtein screening mechanism works in the context of DHOST theories, one considers a non-relativistic spherical object of density $\rho(r)$, which induces a perturbation of the metric, expressed in terms of the Newtonian potentials $\Phi$ and $\Psi$ as
\begin{equation}
ds^2=-\left[1+2\Phi(r)\right] dt^2+\left[1-2\Psi(r)\right]\delta_{ij}\, dx^i dx^j\,.
\end{equation}
The spherical object also perturbs the scalar field,
\begin{equation}
\phi=\phi_c(t)+\chi(r)\,,    
\end{equation}
and the mass inside the radius $r$ is
\begin{equation}
    m(r)=4\pi\int_0^r \bar{r}^2 \rho(\bar{r})d\bar{r} \, .
\end{equation}
The two Newtonian potentials satisfy the differential equations \cite{Dima:2017pwp}
\begin{align}
\label{potentials}
\Phi' (r) &=G_{\rm N} \left( \frac{ m(r)}{r^2} + \Xi_1 { m''(r)} \right)\,, \\
\Psi' (r) & =G_{\rm N} \left( \frac{ m(r)}{r^2} + \Xi_2 \frac{  m'(r)}{ r} + \Xi_3 {  m''(r)} \right)\,,
\end{align}
where
\begin{align}
\label{Upsgen}
\Xi_1 & \equiv \frac{(\alpha_H + c_T^2 \beta_1)^2}{c_T^2 (1 + \alpha_V - 4 \beta_1 ) -\alpha_H -1 } \,, \\
\Xi_2 & \equiv - \frac{\alpha_H (\alpha_H - \alpha_V +2 (1+ c_T^2 ) \beta_1) + \beta_1  (c_T^2-1) (1+c_T^2 \beta_1)}{c_T^2 (1 + \alpha_V - 4 \beta_1 ) -\alpha_H  -1} \,, \\
\Xi_3 & \equiv - \frac{\beta_1 (\alpha_H + c_T^2 \beta_1)}{c_T^2 (1 + \alpha_V - 4 \beta_1 ) -\alpha_H -1 } \,.
\end{align}
The parameter $\alpha_V$ quantifies the non-linear action \cite{Cusin:2017mzw} and is defined in terms of the functions appearing in the action as
\begin{equation}
\alpha_V = {4 X ( F_{2,X} - 2A_2 - 2X A_{2,X} )}/{M^2} \,.
\end{equation}
These coefficients quantify the breaking of the Vainshtein screening inside matter and depend on the DHOST parameters.

Outside the source, $m' = m'' =0$ and hence the standard Newtonian potential is recovered, $\Phi = \Psi = G_{\rm N} m/r$. Inside the source, the behaviour is modified, similarly to the beyond-Horndeski scenario  \cite{Kobayashi:2014ida}.

The screening in DHOST theories has been studied in the context of the  LIGO/Virgo observations in \cite{Crisostomi:2019yfo}, and stringent constraints on the $\beta_1$ parameter are placed. Mechanisms to evade these gravitational waves constraints have been proposed in \cite{Amendola:2018ltt, Frusciante:2018tvu, Hirano:2019scf}.

\subsection{Newtonian and Neutron Stars}
These results can be used to investigate the impact of such theories on various astrophysical bodies. In the case of Newtonian stars, there is a theoretical lower bound for the function $\Xi_1$  \cite{Saito:2015fza},
\begin{equation}
\Xi_1>-\frac{1}{6} \,.
\end{equation}
This quantity can also be constrained from observations, including the lowest mass hydrogen burning stars \cite{Sakstein:2015zoa}, white dwarfs \cite{Saltas:2018mxc} and by analysing weak lensing and X-ray profiles of galaxy clusters \cite{Sakstein:2016ggl,Babichev:2016jom}.

Neutron stars have been studied in a subclass of DHOST models, depending on a single free function $F(X)$ \cite{Boumaza:2022abj},
\begin{equation}
S_{\rm grav}=\int d^{4}x\, \sqrt{-g}\left(F\,R-4\frac{F_X}{X}(L_3-L_4)+L_{\rm m}\right)\,.
\label{S}
\end{equation}
For non-rotating neutron stars, the metric is given by

\begin{equation}
ds^{2}=-f(r)dt^{2}+h(r)dr^2+r^2\left(d\theta^2+\sin^2\!\theta\,  d{\phi}^2\right).
\label{ds}
\end{equation}
Assuming a spherically symmetric scalar field solution of the form \cite{Babichev:2013cya}
\begin{equation}
  \label{phi}
\varphi(t,r)=q\;t+\psi(r)\,,
\end{equation}
the action \eqref{S} becomes
\begin{equation}
S=\int d^{4} x \, r^2\sqrt{f h}\left(\frac{2 F h'}{h^2 r}+\frac{2 F (h-1)}{h r^2}-\frac{4 F_X X' \left(f X+q^2\right)}{f h r X}+L_m\right)\,,
\label{S2}
\end{equation}
where
\begin{equation}
\label{X}
X=\psi'^2/h-q^2/f\,.
\end{equation}
Here the matter appearing in the Lagrangian $L_m$ is modelled  by a perfect fluid with energy density $\rho$, pressure $P$ and 4-velocity $u^\mu$. The  energy-momentum tensor is thus  
\begin{equation}
T^{\mu\nu}=\frac{2}{\sqrt{-g}}\frac{\delta (\sqrt{-g}\,  L_m)}{\delta g_{\mu\nu}}=(\rho+P)u^\mu u^\nu + Pg^{\mu\nu}\,.
\end{equation}
From the energy-momentum tensor one obtains
\begin{eqnarray}
P'=-\frac{f'(P+\rho)}{2f}\,.
\label{e4}
\end{eqnarray}
Considering a Taylor expansion of the function $F$ of the form,
\begin{equation}
    F=\frac{\kappa}{2}+\sigma X\,,
\end{equation}
the action \eqref{S2} can be varied to find solutions for $\psi$. 
Then redefining $\psi$ as
\begin{equation}
\psi \to \frac{p^{1/2}}{q} \psi\,, \qquad p\equiv  q^2 \sigma\,,
\end{equation}
one gets
\begin{eqnarray}
\psi'^2 =\Delta_1-\Delta_2,\label{ex}
\end{eqnarray}
where
\begin{eqnarray}
\Delta_1 &=& \frac{h \left[f \left(2 \kappa + r^2 P\right)+8 p \right]-4 p}{8 f} \, ,\\
\Delta_2^2 &=& \Delta_1^2-\frac{h p \left[f \left(h \left(2 \kappa +2 P r^2+ \rho  r^2\right)-2 \kappa \right)+4 p
   \left(h -1\right)\right]}{4 f^2} \,.
\end{eqnarray}
The deviations from GR $\Xi_1$ and $\Xi_2$ are given by \cite{Langlois:2017dyl}
\begin{eqnarray}
\Xi_1 =-2\, \Xi_2 = \frac{2p}{2p+\kappa}.
\end{eqnarray}
Outside the star $X=-q^2$ and hence the modification is indistinguishable from GR.
Inside the star, their radial profiles must be considered, which depend on their central density and on the equation of state. This can be achieved by numerically integrating the equations of motion. In \cite{Boumaza:2022abj} several scenarios have been considered and it was shown that in DHOST theories neutron stars can have masses and radii significantly larger than in GR, of up to 50\% depending on the numerical values considered for the parameters, which could provide explanations for several observations, such as the pulsar PSR J1614-2230 \cite{Demorest:2010bx} or the compact object measured by GW190814 event \cite{LIGOScientific:2020zkf}, which have these properties.

These results were extended to slowly rotating stars \cite{Hartle:1967he,Hartle:1968si}. The metric \eqref{ds} is generalised to
\begin{eqnarray}
ds^{2}=-f(r)dt^{2}+h(r)dr^2+r^2\left(d\theta^2+\sin^2\!\theta\,  d{\phi}^2\right)
+2 w(r, \theta) r^2 \sin^2\!\theta\,  dt d\phi \, .\label{ds2}
\end{eqnarray}
The function $w$, related to rotation, is assumed to be small, i.e. $w(r, \theta)\ll f$. The perfect fluid in the star is also rotating, which can be described by the four-velocity
\begin{eqnarray}
u^\mu = \frac{1}{f^{1/2}}\{\;1\;,\;0\;,\;0\;,\;\Omega\; \}+{\cal O}\left(\Omega^2 \right)\,,
\end{eqnarray}
up to first order in the fluid angular velocity $\Omega=d\phi/dt=u^\phi/u^t$. The second order action analogous to \eqref{S2} is
\begin{eqnarray}
S^{(2)}=\int d^4 x \sin^3\theta \left(\frac{F r^4 }{2 \sqrt{f} \sqrt{h}} (\partial_r w)^2+\frac{F \sqrt{h} r^2 }{2 \sqrt{f}}(\partial_\theta w)^2+\frac{\sqrt{h} \rho  r^4 }{2 \sqrt{f}}w^2+L_m^{(2)}\right)\,,
\end{eqnarray} 
which yields the equations of motion
\begin{align}
\partial_{rr} w+& \left(\frac{4}{r}-\frac{f'}{2 f}-\frac{h'}{2 h}+\frac{F_X X'}{F}\right) \partial_{r}w+\frac{h}{r^2} \left(\partial_{\theta\theta} w+3 \cot\theta\, \partial_{\theta}w\right)\nonumber\\
&-\frac{h   (P+\rho )}{F}(\Omega+w)=0\,.
\label{eq_omega}
\end{align}
These equations can be expressed as an infinite sum of Legendre polynomials
\begin{eqnarray}
w(r,\theta)= \sum_{l} w_l (r) \, P_{l}(\theta)\,.
\label{omegalegndre}
\end{eqnarray}
with $w_l$ satisfying
\begin{eqnarray}
&&\frac{(hf)^{\frac{1}{2}}}{r^4 F}\frac{d}{dr}\left[\frac{r^4 F}{(hf)^{\frac{1}{2}}}w_l'\right]+\frac{h ((l+1)l-2)}{r^2} w_l-\frac{h   (P+\rho )}{F}(\Omega+w_l)=0 \, .\label{omega}
\label{eq_w_l}
\end{eqnarray}
Outside the star, the results are the same as in GR. Inside the star, the equation can be solved numerically for each $l$. 
For $l=1$ one gets,
\begin{equation}
w_1''+\frac4r w_1'=0\,,
\end{equation}
which has a solution of the form 
\begin{eqnarray}
w_1=-2\frac{J}{r^3} \,.
\end{eqnarray}
The integration constant $J$ physically corresponds to the angular momentum. The moment of inertia is  defined as $I= J/\Omega$. 
Then one can fix the integration constant $J$ in order to match the solutions inside and outside the neutron star. Hence, the moment of inertia of the neutron star $I$ can be expressed as
\begin{eqnarray}
I=\frac{(hf)^{\frac{1}{2}}}{6 F}\int_0^{R}\frac{h^{\frac{1}{2}}(P+\rho )r^4}{f^{\frac{1}{2}}}  \left(1+\frac{w_1}{\Omega}\right)\, dr\,.
\end{eqnarray}
In GR, universal relations relating  the normalised moment of inertia 
\begin{equation}
\tilde I=I/(MR^2)
\end{equation}
 and the stellar compactness
\begin{equation}
{\cal C}=\frac{M}{R}\,,
\end{equation}
as well as the dimensionless moment of inertia 
\begin{equation}
\bar I=I/M^3
\end{equation}
and the compactness ${\cal C}$ \cite{Lattimer:2004nj,Breu:2016ufb} have been derived.
These universal relations are respectively of the form:
\begin{eqnarray}
  \frac{I}{M R^2}&=&a_0+a_1\,  {\cal C}+a_4\,  {\cal C}^4\,,\label{URL}\\
  \frac{I}{M^3}&=&a_{-1}\, {\cal C}^{-1}+a_{-2}\,  {\cal C}^{-2}+a_{-3}\,  {\cal C}^{-3}+a_{-4}\,  {\cal C}^{-4},\label{URNL}
  \end{eqnarray}
where the constants $a_i$ can be obtained numerically.
  
In \cite{Boumaza:2022abj}, the robustness of these universal relations has been analysed in the context of DHOST theories for various deviations from GR, showing that $\mid 1-I/I_{fit}\mid <0.1 / 0.05$ for relations (\ref{URL}) and (\ref{URNL}) respectively.

\subsection{Black Holes}
\label{sec:bh}
A renewed interest in black holes can be attributed to  the LIGO/Virgo direct detection of gravitational waves from binary black hole and neutron star  mergers. Black holes can also be used as probes to investigate deviations from GR. Hence, one has to make sure that theories of modified gravity or dark energy yield constraints compatible with  black holes. In standard GR, the most general axisymmetric stationary rotating black hole solution is the Kerr solution \cite{1968CMaPh..10..280C,CARTER1968399}. Indeed, in the binary black hole observation, it is thought that the merger involved two spinning Kerr black holes \cite{LIGOScientific:2016aoc}. These GR Kerr solutions can be generalised to DHOST theories. In \cite{Charmousis:2019vnf}, DHOST Kerr black hole solutions have been derived,  which have \textit{stealth hair}, i.e.  the scalar field is non-trivial. Linear perturbations around these solutions are discussed in \cite{Charmousis:2019fre}. Methods for generating solutions by using disformal transformations are discussed in \cite{BenAchour:2020wiw, BenAchour:2020fgy}, or by using perturbations about the flat Minkowski metric $\eta_{\mu \nu}$ along a null direction (the Kerr-Schild ansatz) in \cite{Babichev:2020qpr,Baake:2021jzv}. 

We present here the Kerr solutions from DHOST theories, and we mention at the end of the section other recent developments in the field. Quadratic shift-symmetric DHOST theories with $c_T=1$, where all the functions in the Lagrangian only depend on $X$ are considered, such that  $X = X_0$ is a constant. Moreover, stealth solutions are sought (solutions where the metric $g_{\mu \nu}$ is also a solution of the vacuum Einstein equations with a cosmological constant). Although such solutions are plagued by the strong coupling problem \cite{Minamitsuji:2018vuw, deRham:2019gha}, a controlled detuning, the scordatura mechanism \cite{Motohashi:2019ymr, Gorji:2020bfl}, can be employed to solve the issue. The assumptions imply that
\begin{align}
\label{condStealth}
P + 2 \Lambda F = 0 \, , \quad
P_X + \Lambda (4 F_X - X_0 A_{1X}) = 0 \, , \quad
Q_X=0 \, , \quad
A_1 = 0 \, \quad
A_3 + 2 A_{1X} = 0 \, ,
\end{align}
where the functions are evaluated at $X=X_0$.

By considering solutions that satisfy $c_T=1$, one obtains
\begin{align}
\label{cequal1}
&A_1=A_2=0 \, , \qquad Q=0 \, , \nonumber\\
&P(X_0) + 2 \Lambda F(X_0) = 0 \, , \quad
P_X(X_0) + 4 \Lambda F_X(X_0)  = 0 \, , \quad
A_3(X_0) = 0 \, .
\end{align}
The metric that is obtained is the usual Kerr solution in GR, which can be expressed in  Boyer-Lindquist coordinates $(t,r,\theta,\psi)$ as
\begin{align}
ds^2 = - \frac{\Delta_r}{\Xi^2 \rho} \left( dt - a \sin^2{\theta} \, d\psi \right)^2 + \rho \left( \frac{dr^2}{\Delta_r} + \frac{d\theta^2}{\Delta_{\theta}}\right) + \frac{\Delta_{\theta} \sin^2{\theta}}{\Xi^2 \rho} \left( a dt - \left(r^2 + a^2 \right) d\psi \right)^2 \,,
\label{Kerrmetric}
\end{align}
where the cosmological constant $\Lambda=3/\ell^2$, $\Xi \equiv 1 + {a^2}/{\ell^2}$ is a constant, and 
\begin{align}
\Delta_r & = \left( 1 - \frac{r^2}{\ell^2}\right) \left( r^2 + a^2\right) - 2 Mr \;, \qquad  \Delta_{\theta}  = 1 + \frac{a^2}{\ell^2} \cos^2{\theta} \;,\quad \rho  = r^2 + a^2 \cos^2{\theta} \, .
\end{align}
$M$ is the mass of the black hole and $a$ the angular momentum parameter satisfying the condition $a \leq M$.

These stealth black holes also have scalar hair characterised by \cite{Charmousis:2019vnf}
\begin{equation}   
\label{profilekerr}
\phi(t,r,\theta) = -Et + S_r(r) + S_{\theta} (\theta) \, , \qquad
S_r \equiv  \pm \int dr \, \frac{\sqrt{\mathcal{R}}}{\Delta_r} \;, \quad S_{\theta} \equiv \pm \int d\theta \, \frac{\sqrt{\Theta}}{\Delta_{\theta}} \,,
\end{equation}
where $E$ is a constant and
\begin{equation}
    \label{func}
\mathcal{R}(r) \equiv m^2 \left( r^2 + a^2\right) \left[ \eta^2 \left( r^2 + a^2\right) - \Delta_r\right]\;, \qquad \Theta(\theta) \equiv a^2 m^2 \sin^2{\theta} \left( \Delta_{\theta} - \eta^2 \right) \, ,
\end{equation}
with $X_0=-m^2$ and $\eta \equiv \Xi E /m$.
If there is no cosmological constant ($\ell \to \pm \infty$), then $\eta=1$, $\Theta=0$, $E=m$ and
\begin{equation}
    \label{funcR0}
\mathcal{R}(r) \equiv 2 M m^2  r \left( r^2 + a^2\right)\; .
\end{equation}
New Kerr solutions can be derived by considering disformal transformations of the metric of the form
\begin{equation}
\tilde{g}_{\mu\nu} = A(X) g_{\mu\nu} + B(X) \phi_\mu \phi_\nu \, ,
\end{equation}
where $A$ and $B$ are the disformal potentials, depending on $X$. DHOST theories are stable under these transformations \cite{Achour:2016rkg}. Hence, assuming that the tilded action satisfies the stealth conditions (\ref{condStealth}), with a Kerr solution (\ref{profilekerr}), the conditions for the untilded equations become
\begin{align}
&P + \frac{2 \Lambda N}{A_0} F = 0 \, , \qquad
{\partial_X} \left[ P + \frac{\Lambda}{A_0} \left( 4 + \frac{B_0 X_0}{N}\right) F - \frac{\Lambda X_0}{N^3} A_1\right] = 0  \, , \label{cond1a}\\
& Q_X = 0 \, , \qquad A_1 - \frac{N^2 B_0}{A_0} F = 0 \, , \\ 
&\frac{A_0}{2N} A_3 +  \left( 2 B_0 N_X + \frac{B_0 N^2}{A_0}\right)F + 2 B_0 N F_X + B_0 N A_2 + \frac{N^8}{A_0}
\partial_X \left( \frac{A_1}{N^3} - \frac{B_0}{A_0} \frac{F}{N}\right)=0 \, , \label{cond3a}
\end{align}
where the disformal transformation is constant ($A(X) = A(X_0)\equiv A_0$, $B(X)=B(X_0) \equiv B_0$) and the derivatives are evaluated at $X_0$. Imposing (\ref{cequal1}), these simplify to
\begin{align}
&A_1= - A_2 = N^2 \frac{B_0}{A_0} F \, , \qquad Q=0 \, , \\
&\partial_X \left( P + \frac{4 \Lambda}{A_0} F\right)=0 \, , \qquad
\frac{A_0}{2 N B_0} A_3 + \left(2N_X + \frac{N^2}{A_0}(1-NB_0) \right) F + 2N F_X =0 \, .
\end{align}
In this case, the untilded theory admits the disformed Kerr black hole solution. Fixing $A_0=1$, the metric becomes
\begin{equation}
g_{\mu\nu} \; = \; \tilde{g}_{\mu\nu} - B_0 \, \phi_\mu \phi_\nu \, ,    
\end{equation}
which leads to a new Kerr solution. To avoid a pathological behaviour at $r \to \infty$ (non-asymptotically flat metric at infinity), a vanishing cosmological constant is required, implying $\eta=1$, $\Theta=0$ and  $E=m$ as before.
The scalar field solution becomes
\begin{equation}
\label{profilekerrn}
\phi(t,r) = -m t + S_r(r)  \, , \qquad
S_r =  \pm \int dr \, \frac{\sqrt{\mathcal{R}}}{\Delta}_r \, , \qquad
\Delta_r = r^2+a^2-2Mr \; ,     
\end{equation}
and the new solution is
\begin{align}
\label{newsol}
ds^2 &=- \frac{\Delta_r}{\rho} \left( dt - a \sin^2{\theta} \, d\psi \right)^2 + \frac{\rho}{\Delta_r} {dr^2} + \rho \, {d\theta^2} + \frac{\sin^2{\theta}}{\rho} \left(a \, dt - \left(r^2 + a^2 \right) d\psi \right)^2 \nonumber \\
& + \alpha \left( dt  \pm  {\sqrt{2Mr(r^2+a^2)}}/{\Delta_r} \, dr\right)^2 \, ,
\end{align}
with $\alpha \equiv -B_0 m^2$. The inverse of this metric is
\begin{equation}
\label{inversedisf}
g^{\mu\nu} \, = \, \tilde{g}^{\mu\nu} + \frac{\alpha}{m^2(1-\alpha)} \phi^\mu \phi^\nu \, ,
\end{equation}
where $\tilde{g}^{\mu\nu}$ is the inverse Kerr metric while the only non-vanishing components of $\phi^\mu = \tilde{g}^{\mu\nu} \phi_\nu$ 
are
\begin{align}
\phi^t =  \frac{m}{\Delta_r}\left( r^2+a^2 + \frac{2M r a^2 \sin^2\theta}{\rho}\right) \, , \quad
\phi^r =   m \frac{\sqrt{2M r (r^2+a^2)}}{\rho}\,, \quad
\phi^\varphi =m\frac{2aMr }{\Delta_r \rho} \, .
\end{align}
Hence, the disformal transformation of the stealth Kerr solution considered yields a non-stealth exact solution which is parametrised by an additional deformation parameter $\alpha$ which encodes precisely the deviations from GR, in addition to the  mass and angular momentum parameters $(M, a)$ of the Kerr family. The sign ambiguity in \eqref{newsol} can be absorbed by redefining  $t \to \pm t$ and $a \to \pm a$.

At infinity where $r\to+ \infty$, the disformed Kerr metric becomes equivalent to
\begin{align}
ds^2 &\simeq -\left( 1-\frac{2M_1}{r}\right)dt^2 + \left( 1-\frac{2M_2}{r}\right)^{-1} dr^2 + r^2 (d\theta^2 + \sin^2\theta \, d\varphi^2) \nonumber \\
&+ 2\alpha \sqrt{\frac{2M_1}{r}} dr\,dt + {\cal O}\left( \frac{1}{r^2}\right) \, , 
\end{align}
\label{asymptomet}
where 
\begin{equation}
M_1 \equiv \frac{M}{1-\alpha} \, , \quad M_2\equiv(1+\alpha)M \, . 
\end{equation}
This new solution is not circular \cite{Anson:2020trg}, but as it the case with other Kerr black holes, it does not have a physical singularity.

In \cite{Babichev:2024hjf} it was shown that LISA can be in principle used to distinguish the Kerr geometry in DHOST from GR by measuring the deformation parameter  $\alpha$ which quantifies departures from the standard Kerr black hole spacetime through an accumulated effect. This parameter can also be investigated by looking at its effect on the black hole shadow \cite{Long:2020wqj} using the observations from the Event Horizon Telescope \cite{EventHorizonTelescope:2019dse,EventHorizonTelescope:2019ggy}.

A different class of black holes solutions is represented by the non-rotating Schwarzschild ones, which can be obtained from quadratic DHOST theories \cite{Takahashi:2019oxz,deRham:2019gha, Langlois:2021aji}, though some are disfavoured through a perturbation analysis. Perturbations to such black holes are investigated in \cite{Takahashi:2021bml,Nakashi:2022wdg,Mukohyama:2022skk,Noui:2023ksf}. The strong coupling of the solutions in the DHOST theories is discussed \cite{Motohashi:2019ymr, Gorji:2020bfl}, together with a solution -- adding small corrections (\textit{scordatura}) to avoid this issue and the implication for the black hole solutions \cite{DeFelice:2022qaz}. Various generalisations have been considered, \textit{e.g.}  non-Schwarzschild solutions with a non-constant kinetic term \cite{Minamitsuji:2019tet}.

Black hole solutions have been studied \cite{Minamitsuji:2019shy} for shift-symmetric DHOST theories which admit exact black hole solutions with a linearly time-dependent scalar field. In this case, purely quadratic theories are viable and  exact static and spherically symmetric solutions (Schwarzschild and Schwarzschild-(anti-)de Sitter) are determined. It was also shown that only a subset of the cubic theories admit stealth solutions.

In \cite{Creminelli:2020lxn, Capuano:2023yyh} the no-hair theorem \cite{Misner:1973prb} is generalised to DHOST theories, showing that the coupling with Gauss-Bonnet is necessary to have hair.

\section{Other Constraints}
\label{sec:etc}
In this section we briefly introduce some of the additional effects of DHOST theories that have been studied and that can lead to observable effects.

\subsection*{Sound Speed Constraints}
The modification of the Newtonian potential in DHOST theories with respect to GR induces modifications to the matter perturbations. In \cite{Babichev:2018rfj} it has been shown that these modifications can change the speed of sound in the atmosphere of the Earth and can generate instabilities in homogeneous media. It was shown that this modification can be encoded in a single parameter that can, in principle, be constrained.

\subsection*{Galaxy Cluster Constraints}
DHOST theories modify the Newtonian potential (see Sec. \ref{sec:vain}) and are expected to also induce deviations from GR at galaxy clusters scales hence affecting weak
lensing, X-ray and Sunyaev - Zel’dovich observables. In \cite{Cardone:2020rmy}, expressions for the lensing convergence $\kappa$, and the pressure profile $P$ of clusters in the framework of DHOST theories are determined, quantifying how much they deviate from their GR counterparts. Using the galaxy cluster sample of BOXSZ \cite{Shitanishi:2017xct}, it has been shown that combined measurements of $\kappa$, $P$ and of the electron number density, $n_e$, can constrain both the cluster and DHOST parameters.

In Ref. \cite{Haridasu:2021hzq}, the modifications from DHOST theories were confronted to data coming from the X-COP galaxy cluster sample \cite{Eckert:2016bfe}. Using Bayesian  methods, a constraint of  $\Xi_1=-0.030 \pm 0.043$ (95\% confidence level) has been obtained.

\subsection*{Constraints from Large-Scale Structure}
The large-scale structure of the universe has been used to constrain modified gravity models using two and three-point statistics \cite{Sugiyama:2023tes}. In this work a joint analysis of the anisotropic components of galaxy two- and three-point correlation functions (3PCF) to actual galaxy data and has been performed, together with the computation of  constraints of the nonlinear effects of DHOST  theories on cosmological scales. Using the Baryon Oscillation Spectroscopic Survey (BOSS) data release 12,  lower bounds of $-1.655 < \xi_t$ and $-0.504 < \xi_s$ at 95\% confidence level on the parameters characterising the time evolution of the tidal and shift terms of the second-order velocity field were derived. These constraints are consistent with GR predictions of $\xi_t =15/1144$ and $\xi_s =0$. This represents a 35 and 20-fold improvement respectively, over using only the the isotropic 3PCF.

\subsection*{Helioseismology Constraints} 

The screening mechanism described in Sec. \ref{sec:vain} provides an opportunity to perform precision tests of gravity at stellar scales. In \cite{Saltas:2019ius,Saltas:2022ybg}, the Sun is used as a laboratory where imprints of the fifth-force effect on the solar equilibrium structure are investigated. The work details how different solar regions can be used to test gravity using analytic techniques and numerical simulations. It is shown that the fifth force leaves an important signature on the solar sound speed, even when accounting for solar microphysics such as opacity, diffusion, equation of state and metallicity. The fifth-force coupling strength of has been constrained to $-10^{-3} \le \mathcal{Y} \le 5 \times 10^{-4}$  (at 95\% confidence level), improving previous results by a factor of three.

\subsection*{Matter Power Spectrum}

In \cite{Hirano:2020dom}, matter density perturbations up to third order and the one-loop matter power spectrum in DHOST theories have been investigated. This has involved systematically solving gravitational field equations and fluid equations order by order. Three shape functions that characterise the third-order solutions in DHOST theories are presented. Convergence conditions of the loop integrals in the infrared
and ultraviolet limits are discussed, showing that these constraints are more stringent than in GR. Hence, the one-loop matter power spectrum is sensitive to the short-wavelength behaviour of the linear power spectrum.

\section{Inflation}
\label{sec:inflation}
The inflationary paradigm, initially designed solve several cosmological paradoxes, including the horizon problem, the flatness problem and the magnetic monopole problem, is also responsible for the formation of the structures through the growth of primordial quantum fluctuations.  Cosmic Microwave background (CMB) experiments, such as COBE \cite{Fixsen:1996nj}, WMAP \cite{WMAP:2012fli} and \textit{Planck} \cite{Akrami:2018odb} have tightened the constraints on the observable effects on inflation. The tightest constraints are based on measurements of the angular power spectrum of CMB anisotropies. The measurements show that the power spectrum  exhibits a small departure from scale invariance that can be quantified through 
\begin{equation}
\mathcal{P}_{\zeta} (k) =  \mathcal{P}_{\zeta} (k_*)  \left(\frac{k}{k_*}\right)^{n_s-1} \, ,
\end{equation}
where $k_*$ is a pivot scale and $n_s$ is the scalar spectral index. Further constraints can be determined by expanding this expression further as
\begin{equation}
\log \mathcal{P}_{\zeta} (k) =  \log \mathcal{P}_{\zeta} (k_*) 
+ \frac{1}{2} \frac{d \log n_s}{d \log k} \left(\frac{k}{k_*}\right)^2 
+ \frac{1}{6} \frac{d^2 \log n_s}{d \log k^2} \left(\frac{k}{k_*}\right)^3 + \ldots \, .
\end{equation}
\textit{Planck} has placed stringent constrains on these coefficients,
\begin{align}
\label{ns_const}
n_s &= 0.9625 \pm 0.0048    \, ,\\
\alpha_s &= \frac{d n_s}{d \ln k} = 0.002 \pm 0.010 \, , \\
\beta_s & = \frac{d \alpha_s}{d \ln k} = 0.010 \pm 0.013 \, , \\
\ln(10^{10} A_s) & = 3.044 \pm 0.014 \,
\label{bs_const}
\end{align}
at $68\%$ confidence level, using the TT,TE,EE+lowE+lensing likelihoods \cite{Akrami:2018odb} at a pivot scale of $k_* = 0.05 \mathrm{~Mpc}^{-1}$.

An additional constraint comes from the tensor-to-scalar ratio, currently bounded at the level of $r < 0.044$ at 95\% confidence \cite{Tristram:2020wbi} when combining \textit{Planck} with BICEP2/Keck 2015 data \cite{Ade:2018gkx}. 

Further constraints can be determined by investigating predictions on higher-order correlators, such as the bispectrum (three-point correlation function). In particular, \textit{Planck} has placed the tightest constraints so far on the amplitudes of non-Gaussianities, which are three-dimensional quantities. Based on the model of inflation considered, they come predominantly in the following three shapes: local, equilateral and orthogonal, which can be expressed in terms of the primordial power spectrum $P_{\Phi}$ as
\begin{align}
\label{eq:Bloc}
B_{\Phi}^{\text{loc}}&(k_1,k_2,k_3)=2 \left[P_{\Phi}(k_1)P_{\Phi}(k_2)+\text{2 perms} \right] \, ,\\
B_{\Phi}^{\text{equil}}&(k_1,k_2,k_3)=6 \left\{-[P_{\Phi}(k_1)P_{\Phi}(k_2)+\text{2 perms} ]\right. \nonumber \\
&-2[P_{\Phi}(k_1)P_{\Phi}(k_2)P_{\Phi}(k_3)]^{2/3} 
+[P_{\Phi}^{1/3}(k_1)P_{\Phi}^{2/3}(k_2)P_{\Phi}(k_3)+\text{5 perms}]\left. \right\} \,, \\ 
  B_{\Phi}^{\text{orth}}&(k_1,k_2,k_3) = 6\big[ 3(P_\Phi^{1/3}(k_1)P_\Phi^{2/3}(k_2)P_\Phi(k_3)+5\text{ perms}) \nonumber  \\
  &-3 \left[P_{\Phi}(k_1)P_{\Phi}(k_2)+\text{2 perms} \right] 
  -8 (P_\Phi(k_1)P_\Phi(k_2)P_\Phi(k_3))^{2/3}
\big] \,.
\label{eq:Borth}
\end{align} 
For a bispectrum of one of these shapes,  $f_{\rm NL}$ represents the amplitude of its non-Gaussianities. The current \textit{Planck} constraints on these shapes are 
\begin{align}
\label{eq:fnlloc}
-11.1<f_{\rm NL}^{\rm local} &<9.3 \,,\\
-120<f_{\rm NL}^{\rm equil} &< 68 \,,\\
-86<f_{\rm NL}^{\rm orth} &<10 \,, 
\label{eq:limits}
\end{align}
at 95 \% CL \cite{Akrami:2019izv}.

As mentioned in the introduction, theories designed to explain dark energy can also be adequate to describe inflation. In the case of DHOST theories, work has been performed to build models compatible with the constraints on inflation from both power spectrum \cite{Brax:2021qlx} and bispectrum \cite{Brax:2021bok, Sohn:2023dbp,Lazanu:2024wnk}. 

The starting point is a simplification of the quadratic DHOST action,
\begin{eqnarray}\label{eq:action-dhost-inf}
S_{\rm D} = \int d^4 x \sqrt{-g} \bigg[ F_0(X) + F_1(X) \Box\phi + F_2(X) R
+ \frac{6 F_{2,X}^2}{F_2} \phi^{\nu} \phi_{\nu\eta} \phi^{\eta\lambda}\phi_{\lambda} \bigg]\,,
\end{eqnarray}
where an assumption has been made that the functions the $F_i$ and $A_i$ only depend on the kinetic term $X$.

This action can be used to determine the power spectrum of primordial curvature perturbations in the DHOST theories, using the methods developed in \cite{Gorji:2020bfl}.

It turns out that inflationary solutions in a de Sitter phase of the universe would lead either to infinite strong couplings or gradient instability for scalar field perturbations. This problem can be rectified by slightly breaking the degeneracy conditions, adding \textit{scordatura} corrections of the form $\alpha L_i^{(2)}$, for one $i=1,\dots,5$ (see Sec. \ref{sec:bh}). In this case, the theory is shift-symmetric (invariant under the symmetry $\phi \to \phi + c$). It can be shown that such theories lead to scale-invariant power spectra, that are not compatible with observations ($n_s=1$ is not allowed by CMB data).

To  resolve this issue, one can consider Eq. (\ref{eq:action-dhost-inf}) as the background and add an axion-like potential \cite{Marsh:2015xka}  as a perturbation 
\begin{equation}
\label{eq:lambdam2}
   S_{\rm V} = -\int d^4  x \sqrt{- g}  \mu^4 \left(\cos \frac{\phi}{f}-1\right) \approx \int d^4 x \sqrt{-g} \bigg[ - \frac{m_{\rm phys}^2}{2}  \phi^2 - \frac{\lambda_{\rm phys}}{4!}  \phi^4 \bigg] \,,
\end{equation}
where
\begin{equation}
m^2_{\rm phys}=- \frac{\mu^4}{f^2},\ \lambda_{\rm phys}= \frac{\mu^4}{f^4} \, ,
\end{equation}
for $\phi \ll f$.

The determination of the power spectrum has been performed in \cite{Brax:2021qlx}, based on \cite{Gorji:2020bfl}, following the standard formalism of inflationary perturbation theory \cite{Peter:2013avv}. Here we illustrate the procedure for the Lagrangian without the perturbations $m^2$ and $\lambda$, with the full details given in \cite{Brax:2021qlx}. It is convenient to work with dimensionless coordinates (tilded) and variables (expressed in lowercase), which can be achieved by using two mass scales, $M$ and $\Lambda \simeq m_{\rm Pl}$, defined by:
\begin{align}
&{\tilde t} \equiv \Lambda t \,, \hspace{1cm} {\tilde x}^i \equiv \Lambda x^i \,, \\
\phi \equiv M \, \varphi\,, &\hspace{.5cm} X\equiv{M^2 \Lambda^2}{\mathrm x}\,, \hspace{.5cm} 
F_0 \equiv \Lambda^4 f_0 \,, \hspace{.5cm}
F_1 \equiv \frac{\Lambda^2}{M} f_1 \,, \hspace{.5cm} F_2 \equiv \Lambda^2 f_2 \,.
\end{align}
The action is perturbed at first order in perturbations and expanded up to quadratic order, and then, after suitable integrations by parts, the second order DHOST Lagrangian is obtained in terms of the comoving curvature perturbation $\zeta$, 
\begin{equation}
\label{eq:l2}
{\tilde {\cal L}}_{\rm D}^{(2)}
= a^3 f_2 \bigg( \bar{\cal A} \, \dot{\zeta}^2 - \bar{\cal B} \, \frac{{\tilde k}^2}{a^2} \zeta^2
\bigg) \,,
\end{equation}
with 
\begin{eqnarray}\label{coefficients-AD-BD}
\bar{\cal A} = 6 \bigg[ 1 - \frac{\beta_K}{(1+\alpha_B)^2} \bigg] \,, \hspace{1cm}
 \bar{\cal B} = - 2 \bigg[ 1 - \frac{1}{a f_2} \frac{d}{d{\tilde t}} 
\bigg(\frac{af_2}{h_b}\frac{1+\alpha_H}{1+\alpha_B}\bigg) \bigg] \,.
\end{eqnarray}
Here the parameters $\alpha_i$ and $\beta_i$ depend on the functions $f_i$ and their derivatives and are given by
\begin{equation}\label{alpha-i}
\alpha_H \equiv - {\rm x} \frac{f_{2,{\rm x}}}{f_2}\,, \hspace{1cm}
\alpha_B \equiv \frac{1}{2} \frac{\dot{\varphi}\,{\rm x}}{h_b} \frac{f_{1,{\rm x}}}{f_2} + \alpha_H \,, \hspace{1cm}
\alpha_K \equiv - \frac{{\rm x}}{6h_b^2}\frac{f_{0,{\rm x}}}{f_2} + \alpha_H + \alpha_B \,,
\end{equation}
\begin{eqnarray}\label{beta-K}
&&\beta_K \equiv - \frac{{\rm x}^2}{3} \frac{f_{0,{\rm x}{\rm x}} }{h_b^2 f_2} 
+ (1-\alpha_H) (1+3 \alpha_B) + \beta_B 
+ \frac{(1 + 6 \alpha_H - 3 \alpha_H^2) \alpha_K
- 2 ( 2 - 6 \alpha_H + 3 \alpha_K ) \beta_H}{1-3 \alpha_H} \,,
\nonumber \\
&& \beta_B \equiv  \dot{\varphi}\,{\rm x}^2 \frac{f_{1,{\rm x}{\rm x}}}{h_b f_2} \,,
\hspace{1cm}
\beta_H \equiv {\rm x}^2 \frac{f_{2,{\rm x}{\rm x}}}{ f_2}
\,.
\end{eqnarray}

This second-order action can be used to derive the power spectrum through field quantisation \cite{Sasaki:1983kd, Kodama:1984ziu, Mukhanov:1988jd}.
To examine the creation of primordial fluctuations from the Bunch-Davies vacuum in these models, it is convenient to parameterise the de Sitter phase in the early universe. Hence, the Lagrangian \eqref{eq:l2} can be used to express the second-order action as 
\begin{equation}
\label{eq:S2p}	
	S_{2,s} = \int d\tau d^3k z^2 \left[ \zeta^{\prime 2} - c_s^2 k^2 \zeta^2 \right] \,,
\end{equation}
where primes denote derivative with respect to conformal time; the Mukhanov-Sasaki variable $v_k$, defined as $v=z\zeta$, satisfies

\begin{equation}
	\label{eq:MSscalar}
	v_k^{\prime \prime} + \left(k^2c_s^2 - \frac{z^{\prime \prime}}{z}\right) v_k = 0 \,,
\end{equation}
This equation can be solved using appropriate conditions at $\eta \to - \infty$ and the power spectrum of (scalar) curvature perturbations is computed as 
\begin{equation}
\label{eq:Pzeta}	
	P_{\zeta}(k) = \frac{k^3}{2\pi^2} \frac{|v_k(\eta_E)|^2}{z(\eta_E)^2} \, ,
\end{equation}
where $\eta_E$ is the time corresponding to the horizon entry.

During the inflationary epoch, as solutions linear in time are sought, $X=-1$  and hence the $\alpha_i$ and $\beta_i$ functions become constants. There are five constant functions (and an additional two perturbation parameters) that  can be chosen to match the CMB observations. In \cite{Brax:2021qlx}, two parameter sets are found. Both sets satisfy the \textit{Planck} constraints, but have different tensor-to-scalar ratios, showing that one would be able to use DHOST theories to describe inflation even if the tensor-to-scalar ratio would be significantly more accurately measured in the future with CMB experiments such as LiteBIRD \cite{Hazumi:2019lys}. These are given by

\begin{align}
\label{eq:infl:b1}
\alpha_B = 1,& \quad \alpha_H = 1.04, \quad \beta_K = 3.97343, \quad f_2=2.7, \quad h_b = 3 \times 10^{-5} \nonumber \\
 & \lambda =10^{-36}, \quad m^2 = -1.6 \times 10^{-23} \quad (r=0.04) \,; \\
\alpha_B = 1,& \quad \alpha_H = 1.001, \quad \beta_K = 3.9993, \quad f_2=8.8, \quad h_b = 10^{-5} \quad \nonumber \\
 & \lambda = 5 \times 10^{-32}, \quad m^2 = -1.5 \times 10^{-27} \quad (r=10^{-3}) \,.
\label{eq:infl:b2}
\end{align}

In the case of the bispectrum \cite{Brax:2021bok}, one can employ the formalism described in \cite{Maldacena2003, Chen:2010xka}: expanding the action at third order in the comoving gauge, computing the Hamiltonian in the interaction picture and using the in-in formalism to determine the three-point correlation function. In addition to the parameters fit for the power spectrum, there are another five that appear in the bispectrum. Due to its 3D nature, the analysis of bispectrum is more involved. As a first order approximation, one can compare the bispectrum of the DHOST model with the three most commonly used templates from \textit{Planck} -- local, equilateral and orthogonal, described in Eqs. \eqref{eq:Bloc}-\eqref{eq:Borth} in the equilateral configuration and take into account the projection of the DHOST bispectrum on these shapes \cite{Babich:2004gb}. This analysis shows that the additional parameters can be tuned such that the overlap between the bispectra and the three templates can be made arbitrarily small \cite{Brax:2021bok}. This result is, however, misleading, as a full likelihood analysis using the \textit{Planck} 2018 CMB results has shown that the bispectrum consists of a fixed contribution (coming from the power spectrum parameters) and a linear combination of terms depending on five new parameters defining the cubic perturbations. They have different shapes, with the former peaking in the squeezed limit, and the latter in the equilateral limit. These shapes cannot cancel sufficiently if the amplitude of the squeezed shape is too large. It has thus been shown in \cite{Sohn:2023dbp} that the model described in Eq. \eqref{eq:infl:b1} is not favoured by the data. By using the CMB-BEST code \cite{Sohn:2023fte} and marginalising over the free parameters it has been shown explicitly that there are viable DHOST inflationary models satisfying both power spectrum and bispectrum constraints from \textit{Planck}, with a model satisfying both power spectrum and bispectrum constraints having parameters
\begin{align}
\alpha_B = 1,& \quad \alpha_H = 1.04, \quad \beta_K = 3.97343, \quad f_2=6, \quad h_b = 0.00001712, \nonumber \\
& \lambda = 1.2 \times  10^{-36}, \quad m^2 = - 10^{-23}  \,.   
\end{align}

\section{Other Developments for the Early Universe}
\label{sec:bounce}
The initial singularity problem appearing in inflation has led to the development of alternative theories. Such alternatives include bouncing cosmologies \cite{Gasperini:1992em} and Galilean Genesis models. The instability problem in bouncing cosmology has been studied in Refs. \cite{Libanov:2016kfc, Kobayashi:2016xpl, Cai:2016thi,Cai:2017tku, Cai:2017dyi, Creminelli:2016zwa}. In \cite{Cai:2016thi}, it was found for the first time that a stable bouncing cosmology can be constructed in a beyond Horndeski theory without encountering a strong coupling problem. These works have led to further investigation of alternatives to inflation within beyond Horndeski or DHOST theories.

In \cite{Ilyas:2020qja,Ilyas:2020zcb,Zhu:2021whu,Zhu:2021ggm} the DHOST Genesis model, inspired by Galilean Genesis \cite{Creminelli:2010ba,Creminelli:2012my} is built using DHOST theories. Potential problems such as no-go theorems for bouncing cosmology \cite{Novello:2008ra,Lehners:2008vx,Battefeld:2014uga,Brandenberger:2016vhg} have been resolved by using DHOST theories \cite{Ilyas:2020qja,Zhu:2021whu}. Similarly, it is shown in \cite{Ilyas:2020zcb, Zhu:2021ggm} that DHOST theories can also evade the no-go theorems for Galilean Genesis cosmology \cite{Creminelli:2010ba, Creminelli:2012my}.  

In this model, the starting action is
\begin{equation}
	\label{eq:action}	
	S = \int d^4x \sqrt{-g} \left( \frac{R}{2} + \mathcal{L}_{H2} + \mathcal{L}_{H3} + \mathcal{L}_D \right) \,,
\end{equation}
where $\mathcal{L}_{H2}$, $\mathcal{L}_{H3}$ and $\mathcal{L}_D$ are given by 
\begin{align}
	\mathcal{L}_{H2} &= K(\phi,X) = - g_1(\phi) X + g_2(\phi) X^2 \,,  \\
	\mathcal{L}_{H3} &= G(X)\Box X ~,~ G(X) = \gamma X^2 \,,  \\
	\label{eq:DHOST}
	\mathcal{L}_D & =  \frac{R}{2}h - \frac{h}{4X} \left[ \phi_{\mu \nu} \phi^{\mu \nu} - (\Box \phi)^2 \right] 
	+ \frac{h-2Xh_X}{4X^2} \left[ \phi_{\mu}\phi^{\mu \rho}\phi_{\rho \nu}\phi^{\nu} - (\Box \phi)\phi^{\mu}\phi_{\mu \nu}\phi^{\nu} \right] \, ,
\end{align}
where 
\begin{equation}
	\label{eq:g12}
	g_1(\phi) = \frac{3}{2}f^2 e^{4\phi} \frac{1+e^{2\phi}}{1 + e^{m \phi}} \,,~ g_2(\phi) = e^{2\phi} \frac{1 + e^{2\phi}}{1+e^{4\phi}} \,, 
\end{equation}
\begin{equation}
	h(X) = d_1 X + d_2 X^2 \, .
\end{equation}
The parameter $m>6$ determines the behaviour when the universe is exiting the quasi-Minkowskian state. Throughout the analysis, $m=7$ through is used.

For the background, the Friedmann equations are obtained,
\begin{equation}
\label{eq:Friedmann}	
	3H^2 = \rho_{\phi} ~,~ -2\dot{H} = \rho_{\phi} + p_{\phi} \,,
\end{equation}
where $\rho_{\phi}$ and $p_{\phi}$ are the density and pressure of the matter field $\phi$, expressed by
\begin{equation}
	\label{eq:rho}
	\rho_{\phi} = -\frac{1}{2} g_1 \dot{\phi}^2 + \frac{3}{4}g_2 \dot{\phi}^4 + 3H\gamma \dot{\phi}^5 \,,
\end{equation}
\begin{equation}
	\label{eq:p}
	p_{\phi} = -\frac{1}{2}g_1 \dot{\phi}^2 + \frac{1}{4}g_2 \dot{\phi}^4 - \gamma \dot{\phi}^4\ddot{\phi} \, ,
\end{equation}
independent of the DHOST term \eqref{eq:DHOST} \cite{Ilyas:2020qja,Ilyas:2020zcb}.
Assuming as the initial condition of scalar field $\phi$ in the far past to be $\phi \ll -1$, the first Friedmann equation can be written as
\begin{equation}
	3H^2 = -\frac{3}{4} f^2 e^{4\phi} \dot{\phi}^2 + \frac{3}{4}e^{2\phi} \dot{\phi}^4 + 3H\gamma \dot{\phi}^5 \,,
\end{equation}
yielding a solution for the pressure
\begin{equation}
	p_{\phi} = -\frac{1}{(-t)^6} \left( \frac{1}{2f^2} + \gamma \right) \, .
\end{equation}
This shows that the universe will deviate from the quasi-Minkowskian configuration.

The background evolution of the model can be described through the Friedmann equation

\begin{equation}
	-2\dot{H} \simeq p_{\phi} = -\frac{1}{(-t)^6} \left( \frac{1}{2f^2} + \gamma \right) \,,
\end{equation}
where
\begin{equation}
	\label{eq:Ht<0}
    H = \frac{h_0}{(-t)^5} ~,~ h_0 \equiv \frac{1}{10} \left( \frac{1}{2f^2} + \gamma \right) \,.
\end{equation}
Hence,  the pressure and density are described by
\begin{equation}
	\rho_{\phi} = \frac{3}{4}\dot{\phi}^4 + 3H\gamma \dot{\phi}^5 \,,~ p_{\phi} = \frac{1}{4}\dot{\phi}^4 - \gamma \dot{\phi}^4 \ddot{\phi} \,,
\end{equation}
with a radiation-dominated universe solution
\begin{equation}
	a \propto t^{\frac{1}{2}} ~,~ \dot{\phi} \propto t^{-\frac{1}{2}} \,.
\end{equation}
Hence,  the universe starts from a quasi-Minkowskian state, passes through the emergent event at $t = 0$, and finally gets to a standard radiation dominated universe, as in the standard cosmology. 

The power spectrum for scalar and tensor perturbations of the model can be investigated by computing the second order action for linear perturbations, similarly to the standard inflationary scenario (Sec. \ref{sec:inflation}), in terms of the comoving curvature perturbation $\zeta$ and then performing the field quantisation using the Mukhanov-Sasaki prescription,
\begin{equation}
\label{eq:S2s}	
	S_{2,s} = \int d\tau d^3x \frac{z_s^2}{2} \left[ \zeta^{\prime 2} - c_s^2 (\partial_i \zeta)^2 \right] \,,
\end{equation}
where 
\begin{align}
	\frac{z_s^2}{2a^2} &= 3 + 2\frac{\dot{\phi}^2(-g_1 + 3g_2\dot{\phi}^2) + 18H\gamma \dot{\phi}^5 - 6H^2}{(\gamma \dot{\phi}^5 - 2H)^2} \, ,\\
	\left( - \frac{z_s^2}{2a^2} \right) c_s^2 &= 1 + h + \frac{2}{a} \frac{d}{dt} \frac{a \left( h_X \dot{\phi}^2 - h - 1 \right)}{2H - \gamma \dot{\phi}^5} ~.
\end{align}
In the pre-emergent period $t<0$, the universe is almost static, $d\eta = dt/a \simeq dt$ and
\begin{equation}
	\label{eq:zscst<0}
	z_s^2 = \frac{600f^2}{\left(1-8f^2\gamma \right)^2}(-t)^4 ~,~ c_s^2 = \frac{4f^2\gamma-1}{3} \equiv c_{s0}^2 \,,
\end{equation}
and the Mukhanov-Sasaki equation is solved in terms of Bessel functions as
\begin{equation}
	\label{eq:vketa}
	v_k(\eta) \propto \sqrt{k \eta} \left[ c_1(k)J_{\frac{3}{2}}(kc_{s0} \eta) + c_2(k) Y_{\frac{3}{2}} (kc_{s0} \eta) \right] \, .
\end{equation}

The wavemodes of observational interests all become superhorizon in the pre-emergent period $t<0$ \cite{Ilyas:2020zcb}. Hence, the dynamics of scalar perturbation have to be evaluated from past infinity to the critical time scale $\eta_E = t_E$, where the emergent solution breaks down. After $t=t_E$, the universe quickly turns to a radiation dominated epoch, and the scalar perturbation freezes. 

The scalar power spectrum can be evaluated using Eq. \eqref{eq:Pzeta}. By using the vacuum initial conditions it can be easily shown that it is scale invariant as required. 
The tensor power spectrum can be computed analogously and the tensor spectral index is $n_t = \frac{d\ln P_t}{d\ln k} = 2$, consistent with previous studies \cite{Creminelli:2010ba,Nishi:2015pta}.

\vspace{0.7cm}
Another recent contribution regards the influence of higher order terms in the DHOST Lagrangian on non-perturbative solutions, which drastically change the nature of gravitational waves in the non-linear regime, with 
consequences for the early cosmology \cite{BenAchour:2024zzk,BenAchour:2024tqt}. 

Also, a Horndeski/DHOST  phase in the early universe could enhance the production of secondary gravitational waves. This effect could be used to test scalar-tensor theories in the early universe \cite{Domenech:2024drm}.

\section{Conclusions}
\label{sec:conclusions}

In this review we have presented the DHOST theories -- the most general scalar-tensor theories involving one scalar field and its derivatives up to second order, and propagating one scalar degree of freedom and two tensor ones, avoiding the generation of ghosts. They are generalisations of other previously studied theories, and in particular Horndeski theories.

Horndeski theories, first described in 1974, and rediscovered in the 2000s, have long been thought to be the most general scalar-tensor theories having this property because the requirement of having second-order Euler-Lagrange was believed necessary. The building of beyond-Horndeski theories and the understanding of the mechanisms involved in avoiding the propagation of ghosts through degeneracies between the parameters of the models have led the way to the development from the middle of the 2010s of DHOST theories, which include as special cases the previously mentioned ones. Thus, these theories provide a general and flexible framework for parameterising the universe. While a large literature is available covering the less general models, we have not discussed their observational effects in this work in detail, focusing primarily on the DHOST literature.

In recent years, the research on these theories has developed into two directions. Firstly, they have been used to study the late universe and to parameterise dark energy, incorporating previously studied models, exhibiting compatibility with the current cosmological observations and to model astrophysical objects, such as Newtonian and neutron stars as well as black holes. Secondly, they have been applied to the early universe, in the context of its early-time accelerated expansion.

This review covers both these aspects. We discussed the construction of the DHOST theories as generalisations of Horndeski and beyond-Horndeski theories, their classification, observational implications in the late universe, the use of these theories as models of dark energy and cosmological constraints, as well as inflationary model building.

Thus, research related to these theories has initially focused on the late universe, with the aim of understanding current cosmological acceleration, and of testing gravity. Although very general, in order to be compatible with current cosmological and astrophysical observations, the functions appearing in the DHOST theories have to satisfy certain constraints. In particular, the detection of primordial gravitational waves in the GW170817 event, has significantly restricted the parameter space of the models, from both the gravitational wave speed and the avoidance of gravitational waves decay into dark energy. There is still a debate whether these constraints are relevant for cosmological scales, due to the extrapolation of scales involved.

Another avenue of research in DHOST theories which has been active is related to astrophysical objects and in particular to black hole physics. The vicinity of black holes, where strong fields are present, is particularly suited to study deviations from GR, which could be investigated with gravitational wave physics. The different predictions in compact objects of DHOST theories compared to GR can also provide explanations for astrophyical events, such as the one measured by GW190814 event, which has a mass and a radius significantly larger than in GR. This could also be the case for the pulsar PSR J1614-2230. Departures from GR can potentially be measured in neutron stars.

DHOST theories can also be successfully used in the early universe. Models of inflation were built using such theories, and also alternatives to it. The inflationary models have been shown to satisfy the CMB constraints at the power spectrum and bispectrum levels. As the standard model of inflation has an initial Big Bang singularity, alternatives to it have been developed that avoid this issue, such as the bouncing universe scenario, where DHOST theories are well suited. This theories can have distinct phenomenology to standard inflation, and future probes will be able to distinguish between them.

Thus, DHOST theories could be very useful in exploring the fundamental nature of gravity, in fact the motivation for which models of modified gravity have been developed. Future  cosmological and astrophysical probes will be able to increase our understanding of the nature of gravity.

\acknowledgments
This work was supported by a United Kingdom Research and Innovation (UKRI) Future Leaders Fellowship [Grant No. MR/V021974/2]. The author thanks Philippe Brax for the introduction to the topic of DHOST theories and for working together in opening up new aspects of interest.

\section*{Data Access Statement}
No data were created or analysed in this study.

\bibliography{Bibliography}{}

\begin{figure}[!hb]
  \includegraphics[height=5cm]{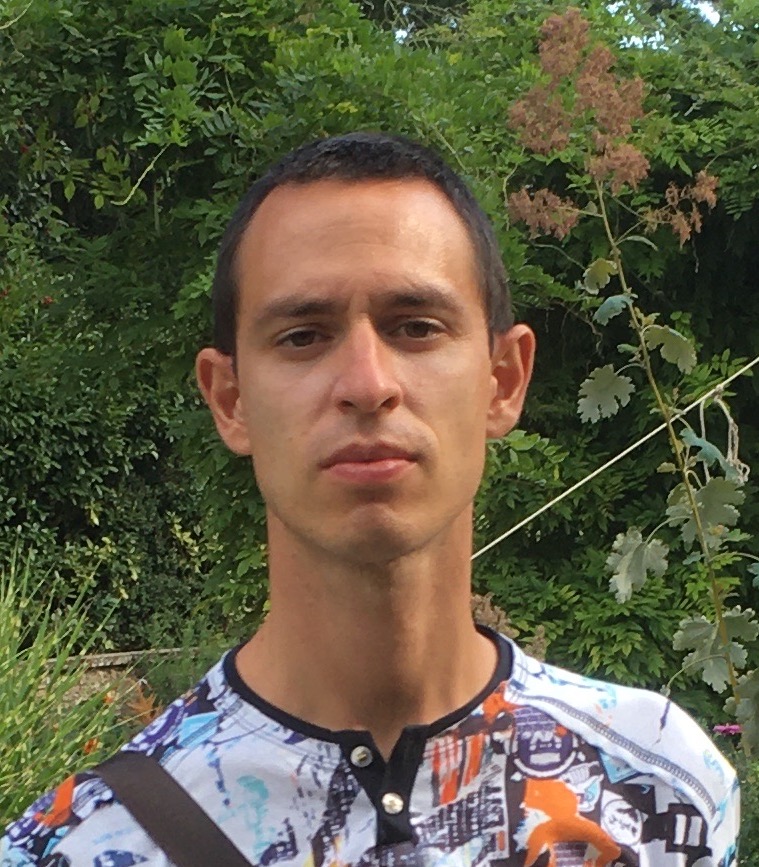}
  \caption*{Andrei Lazanu is a Research Associate at the University of Manchester. He has held positions at ENS Paris and INFN Sezione di Padova. He has been educated at the University of Cambridge (UK), where he has completed the Mathematical Tripos and a PhD in Theoretical cosmology under the supervision of Professor E.P.S. Shellard. AL has a significant international research experience and his interests cover a wide range of topics, including cosmic defects, inflation, the large-scale structure of the Universe, dark energy and the use of machine learning techniques in cosmology.}
\end{figure}

\end{document}